\documentclass[prb,twocolumn,amsmath,amsfonts,10pt]{revtex4-1}

\usepackage{amssymb}
\usepackage[mathscr]{eucal}
\usepackage[dvips]{graphicx}
\usepackage{color}

\newcommand{\ket}[1]{| \, #1 \, \rangle}
\newcommand{\bra}[1]{\langle \, #1 \, |}

\newcommand{\scal}[2]{\bra{#1} \, #2 \, \rangle}
\newcommand{\expect}[1]{\langle #1 \rangle}

\begin{document}

% \title{\fontsize{12pt}{12pt} \selectfont Quantum Chemistry with Tree Tensor Network States}

% \title{\fontsize{12pt}{12pt} \selectfont A novel method to study strongly correlated systems: the Tree-network approach\\ Tree-network based formulation of strongly correlated systems}

\title{\fontsize{12pt}{12pt} \selectfont Simulating Strongly Correlated Quantum Systems with Tree Tensor Networks}

\author{V. Murg}
%\affiliation{Fakult\"at f\"ur Physik, Universit\"at Wien, Boltzmanngasse 3, A-1090 Vienna, Austria}
\author{F. Verstraete}
\affiliation{Fakult\"at f\"ur Physik, Universit\"at Wien, Boltzmanngasse 3, A-1090 Vienna, Austria}
\author{{\"O}. Legeza}
\altaffiliation[Also at ]{Research Institute for Solid-State Physics and Optics, Hungarian Academy of Sciences, H-1121 Budapest, Hungary}
\affiliation{Fachbereich Physik,
  Philipps-Universit\"at Marburg, 35032 Marburg, Germany}
\author{R.\ M.\ Noack}
\affiliation{Fachbereich Physik,
  Philipps-Universit\"at Marburg, 35032 Marburg, Germany}
\date{\today}

\begin{abstract}
We present a tree-tensor-network-based
method to study strongly correlated
systems with nonlocal interactions in higher dimensions.
%RMN characterized by nonlocal interactions
%RMN based on tree tensor networks.
Although the momentum-space
and quantum-chemistry versions of the density matrix renormalization
group (DMRG) method have long been applied to such systems,
the spatial topology of DMRG-based methods allows
efficient optimizations to be carried out with respect to one spatial dimension only.
Extending the matrix-product-state picture, we formulate a more general approach
by allowing the local sites to be coupled to more than two neighboring
auxiliary subspaces.
Following Shi. et. al. [Phys. Rev. A, 74, 022320 (2006)],
we treat a
tree-like network ansatz with arbitrary coordination number $z$, where
the $z=2$ case corresponds to the one-dimensional scheme.
For this ansatz, the long-range correlation deviates from the
mean-field value polynomially with distance, in contrast to the 
matrix-product ansatz, which deviates exponentially.
The computational cost of the tree-tensor-network method is
%RMN considerably
significantly
smaller than that of previous DMRG-based attempts, which 
renormalize several blocks into a single block. 
In addition, we investigate the effect of unitary transformations on
the local basis states and present a method for optimizing such
transformations.
For the 1-d interacting spinless fermion model, the optimized transformation
interpolates smoothly between real space and momentum space.
Calculations carried out on small quantum
chemical systems support our approach.
\end{abstract}

\maketitle

% =============================================

\section{Introduction}

Understanding and simulating strongly correlated systems has long
been a major challenge in theoretical physics and in theoretical chemistry.
In the past two decades,
the density-matrix renormalization group (DMRG) method has been
applied effectively to study problems in these fields.\cite{white92,white92b}
In particular, it
has been widely used to study fermionic
and spin-chain problems in one dimension for models with both local
and long-range interactions.
Application to systems with long-range interactions
gained impetus when the method was reformulated to treat models
defined in momentum space \cite{xiang96,nishimoto02,legeza03} (MS-DMRG) and
quantum chemistry calculations
\cite{white99,daul00,mitrushenkov01,chan02,chan03,legeza03a,legeza03b,moritz05,moritz06,rissler06,marti08} (QC-DMRG).
Common characteristics of these approaches are
that a higher dimensional system is
mapped to a one-dimensional chain and that variational approximations
to the eigenstates are
obtained by an iterative diagonalization procedure.
Introduction of various quantum information entropies
\cite{legeza03,vidallatorre03,vidal03,legeza04} and
the reformulation of the problem in terms of matrix product states
\cite{verstraeteciracmurg08,verstraetecirac05,schuchwolf08}(MPS)
has clarified the mathematical underpinnings of the method.

%In two dimensions, quantum Monte Carlo simulations\cite{ceperley80} have
%long been a powerful approach to study strongly correlated systems.
Recently, extensions to two dimensions 
%However, the fermion sign problem has hindered their application to
%general fermionic problems.
%Recently, 
based on the controlled manipulation of
entanglement between subsystems has led to alternate methods that
can be viewed as generalizations of matrix-product-state-based methods.
%RMN such as the DMRG.
\cite{oestlund95}
These methods include projected entangled pair states
\cite{verstraetecirac04,murgverstraete07,murgverstraete08,verstraeteciracmurg08} (PEPS),
the multiscale-entanglement-renormalization ansatz \cite{vidal06} (MERA),
and correlator product states\cite{changlani09} (CPS) 
or complete-graph tensor network states\cite{marti10} (CGTN).
%RMN While
The PEPS and MERA methods have already shown considerable promise
for frustrated and fermionic problems;
%RMN since 
they do not suffer from the fermion sign problem
that appears 
in quantum Monte Carlo simulations.\cite{ceperley80,troyer04}
These new methods, however, have
primarily been restricted to the treatment of local Hamiltonians.
Thus, developing effective algorithms to treat higher dimensional
systems in which the interactions are nonlocal remains an important problem.

\begin{figure}[b]
    \begin{center}
        \includegraphics[width=0.44\textwidth]{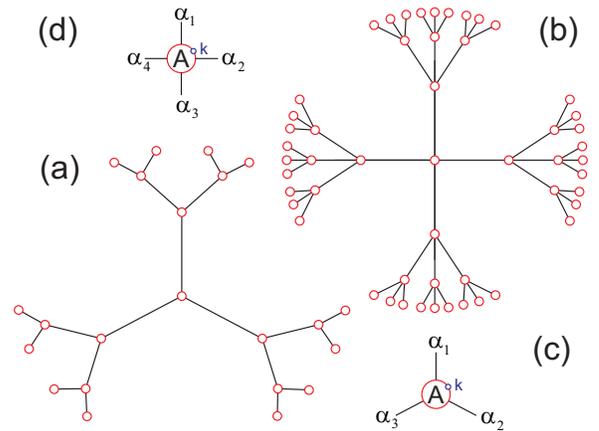}
    \end{center}
    \caption{
        (Color online) Tree tensor network with $z=3$ (a) and $z=4$ (b). The structure of the tensors is shown in (c) and (d). The bonds indicate the virtual indices $\alpha_1,\ldots,\alpha_z$ and the circle the physical index~$k$.
        }
    \label{fig:treenetwork}
\end{figure}

In order to efficiently treat quantum chemical systems, it has become evident in
the past few years that methods must take into account more general
spatial topology.
It has been shown that quantum
information entropy~\cite{legeza03,rissler06} can be used to determine
the entanglement or quantum correlation among sites or orbitals in a
pairwise way.
Such a two-dimensional entanglement matrix leads to a picture of the
topology of orbitals that corresponds to a multiply connected network.
Two possible approaches to optimize computational efficiency are to work with an
appropriate network-like structure that reflects the entanglement
topology and to vary the single-particle basis\cite{luo10} so that entanglement is
reduced.
For states based on a one-dimensional topology, the only possibility
to take entanglement topology
into account is to reorder the orbitals.
The next step in generalizing the topology is to use a 
tree network. 
This was first pointed out %RMN explained 
in Ref.~[\onlinecite{shi06}].
A tree network makes it possible to reduce the distance between
highly entangled orbitals when they are multiply connected.
In addition, the coordination number $z$ can be varied from orbital to
orbital to adapt the variational state to a particular entanglement topology.
For a tree network that is bipartite, it is possible to adapt methods
and optimizations developed for matrix-product-state-based algorithms
such as the DMRG~\cite{schollwoeck04} to
tensor-product states on the tree network.
This is because the tree network method can be related to %RMN an ordinary
a generalized
DMRG with $z$ blocks instead of two. 
In addition, since an effective Hamiltonian can be formed, real time
evolution, 
like that done in t-DMRG\cite{cazalilla02,vidal03,whitefeiguin04,daley04} and
in PEPS\cite{murgverstraete07}, can be carried out.
Real time evolution could be used, for example, to calculate spectral
functions for quantum chemical systems.

The use of more general spatial topologies is also potentially important
for quantum impurity problems. 
In these systems, the impurity subsystem, which can consist of either a single site or
a strongly correlated cluster, is coupled to a bath of free
fermions.
The classical method to treat quantum impurity problems numerically is
the numerical renormalization group (NRG)~\cite{wilson75}.
%%%introduced by K. G. Wilson
The starting point of the NRG method is the mapping of the problem
onto a one-dimensional semi-infinite lattice in which the first site (or
set of sites) describes the impurity subsystem, and %RMN while 
the remainder of
the chain represents the logarithmically discretized conduction band.
In the past few years, the NRG has
been extended by dividing the chain into a system and an environment
as in the DMRG~\cite{hofstetter00};
the resulting method, the density-matrix numerical renormalization
group (DM-NRG), has undergone significant
development recently.\cite{peters06,anders05,toth08,weichselbaum07,holzner10}
While these extensions have led to significant improvements,
they nevertheless have focused primarily on optimizing algorithms
based on a one-dimensional topology.
In general, the conduction bands are
entangled through the impurity subsystem only. 
Therefore, it is a natural choice to describe the problem as a
tree-like structure in which the impurity subsystem at
the center is surrounded by shells of the conduction bath.
Such a tree-like topology has been utilized to treat the quantum
impurity problem that occur with dynamic mean field theory (DMFT).\cite{georges96,eckstein05}

The first attempts to apply matrix-product states on tree networks 
were carried out by Otsuka~\cite{otsuka96}
and by Friedman et. al.~\cite{friedman97}, who calculated ground-state
properties of the spin-1/2 XXZ and Heisenberg chains using the DMRG. %RMN
Subsequently, Lepetit et. al.~\cite{lepetit00} studied the half-filled
Hubbard model on a Bethe lattice (also known as a Cayley tree).
All of this work uses a Bethe lattice, whose characteristics are that the
number of nearest neighbors at each node is $z$, i.e., the coordination
number, and that closed loops do not occur. 
Since there is only one path between any pair
of sites, a DMRG-based solution of the problem is possible.
In this approach, however, $z$ systems blocks must be renormalized to a single
block at each iteration step, leading to  computational
cost which increases exponentially with $z$, hindering systematic
DMRG studies for large systems and/or large $z$; up to now only $z=3$
has been treated.
In contrast, the tree tensor network (TTNS) methods we introduce here
have a much lower computational cost 
because the topology consists of a single site and $z$ blocks,
whereas  the superblock configuration
included $2z$ blocks or ${z+1}$ in previous DMRG attempts. 
%%%In addition, we optimize all the 
%%%$z$ bonds at each iteration step while in previous DMRG attempts only 
%%%one\cite{otsuka} or two bonds\cite{lepetit}
%%% are optimized. 
In this work, we present calculations on small systems in order to demonstrate
the viability of the tree tensor network method and to compare its
accuracy and efficiency with existing DMRG methods.
%%%Like in the early development stage
%%%of the quantum chemistry version of DMRG, 
%%%and do not intend to
%%%provide benchmark results on large molecules.

Another approach to take advantage of the benefits of the tree network
was formulated in Ref.~[\onlinecite{tagliacozzo09}].
In this work, the tree tensor network is formed by placing physical
sites on the boundary sites only. 
The remaining interior sites are virtual; they are only
used to transfer entanglement up the tree.
This tree tensor product state is designed to treat models in which
sites contribute the same amount of entanglement, i.e, have the same
value of single-site entropy. 

In this work, we form a tree tensor
network in which all sites in the tree represent physical sites  
and in which entanglement is transferred via the virtual bonds that
connect the sites.
Our motivation is to treat models in which physical sites have varying
degrees of entanglement; positions closer to the center of the tree
should be better suited to represent more entangled sites.
An additional motivation is to take advantage of the property of the
tree tensor network ansatz
that the long-range correlations differ from the mean-field value 
polynomially with distance rather than
exponentially with distance as for MPS.
In our algorithmic approach to optimize the tree tensor network, we
use tools similar to those used in 
Refs.~[\onlinecite{shi06}] and~[\onlinecite{tagliacozzo09}],
but optimize the network site-by-site as in the DMRG
instead of performing an imaginary time evolution.
In addition, we explicitly describe how to deal
with fermions and long-range interactions.

The paper is organized as follows. 
In Sec.~\ref{sec:theory}, we
describe the theoretical background for the TTNS
algorithm and the orbital optimization used during
the iterative procedure. 
Sec.~\ref{sec:results} is devoted to the analysis
of the error in the ground-state energy for various spin and fermion
models
as a function of bond dimension for TTNS with different coordination numbers.
These include the 2-d Heisenberg model, 
the 2-d spinless fermion model, the 1-d spinless fermion model in momentum space,
and the Beryllium atom as 
an application in quantum chemistry. 
In the latter case, we compare the results %RMN are compared 
to those from DMRG calculations.
We conclude and discuss future prospects in Sec.~\ref{sec:conclusions}.

% =============================================

\section{Theoretical Background of the Tree-Network Approach} \label{sec:theory}

We approach the problem of finding the ground state of strongly correlated systems with long-range interactions  using a TTNS ansatz of the form
\begin{displaymath}
\ket{\Psi} = \sum_{k_1,...,k_M} C_{k_1 \ldots k_M} \ket{k_1,...,k_M}.
\end{displaymath}
Here the coefficients $C_{k_1 \ldots k_M}$ describe a tree tensor
network, i.e., they emerge from contractions of a set of tensors
$\{A_1,\ldots,A_M\}$ according to a tree network, as shown in
Fig.~\ref{fig:treenetwork}. 
%RMN To each vertex~$m$ of the network, a tensor
We associate a tensor with $z+1$ indices, 
\begin{displaymath}
\left[ A_m \right]^{k}_{\alpha_1 \ldots \alpha_z} \, ,
\end{displaymath}
%RMN with $z+1$ indices 
with each vertex~$m$ of the network, that is, each tensor has
$z$ virtual indices $\alpha_1 \ldots \alpha_z$ of dimension~$D$ and one physical index~$k$ of dimension~$d$, with $z$ being the coordination number of that site.
The coefficients $C_{k_1 \ldots k_M}$ are obtained by contracting the
virtual indices of the tensors according to the scheme of a tree
tensor network (see Fig.~\ref{fig:treenetwork}). The structure of the
network can be arbitrary and the coordination number can vary from
site to site. The only condition is that the network is bipartite,
i.e., by cutting one bond, the network separates into two disjoint
parts. 
In the special case $z=2$, the one-dimensional MPS-ansatz used in DMRG
is recovered.

%In the bipartite tree tensor network, each virtual bond connects two disjoint blocks $A$ and $B$. The entanglement between these two blocks is mediated via the virtual bond connecting the blocks. The maximal bond dimension required is equal to the Schmidt-number $\chi$ in the decomposition
%\begin{displaymath}
%\ket{\Psi} = \sum_{\gamma=1}^{\chi} \Gamma_{\gamma} \ket{\Psi_A^{\gamma}} \otimes \ket{\Psi_B^{\gamma}}.
%\end{displaymath}

In a tensor network, entanglement is transferred via the virtual bonds
that connect the sites. Thus, it is preferable to put strongly
correlated sites close together, i.e., to minimize the number of bonds between them. Evidently, the diversity of arranging the sites in the network increases drastically with increasing coordination number~$z$.
Also, with a coordination number $z>2$ the number of virtual bonds
required to connect two arbitrary sites scales logarithmically with
the number of sites~$M$, whereas the scaling is linear in~$M$ for
$z=2$.\cite{shi06}
This can be seen by considering a Cayley-tree of depth~$\Delta$, as shown in Fig.~\ref{fig:treenetwork}. The number of sites in the tree is
\begin{displaymath}
M = 1+z \sum_{j=1}^{\Delta} (z-1)^{j-1} = \frac{z(z-1)^{\Delta}-2}{z-2}
\end{displaymath}
and thus, the maximal distance between two sites, $2 \Delta$, scales logarithmically with~$M$ for $z>2$.
This logarithmic scaling is fundamental because, with a MPS ansatz, the expectation value of a long-range correlation differs from the mean-field result only by a quantity that decays exponentially with the distance:
\begin{displaymath}
\expect{\tau_n \tau_{n+\Delta}} - \expect{\tau_n} \expect{\tau_{n+\Delta}} \propto c^{-\Delta}
\end{displaymath}
With a TTNS ansatz, the logarithmic scaling counteracts this exponential decay, so that the difference from the mean-field result only scales polynomially with the distance. 

The way to arrange the physical sites on the network is determined by
the choice of the basis $\ket{k_1,...,k_M}$. Obviously, the precision
of the ansatz depends critically on the choice of the basis. For
example, the %RMN uncorrelated Fermi-Hubbard model
noninteracting Fermi gas %RMN
has a ground state that is a direct product in the momentum space
representation. %RMN - which
Such a state
corresponds to a tree tensor network state with
virtual dimension $D=1$ at all bonds. In position representation,
however, a bond dimension that increases exponentially with the number
of sites would  be required. Thus, it will be favorable to optimize not only over the tensors in the coefficients $C_{k_1 \ldots k_M}$, but also over the basis $\ket{k_1,...,k_M}$.

%\begin{displaymath}
%C_{k_1 \ldots k_M} =
%\mathcal{F} \big( \big[ A_1
%\big]^{k_1},...,\big[ A_M \big]^{k_M} \big)
%\end{displaymath}

In second quantization, $\ket{k_1,...,k_M}$ denotes the basis in occupation number representation,
\begin{displaymath}
\ket{k_1,\ldots,k_M} =
\left( a_1^{\dagger} \right)^{k_1}
\cdots
\left( a_M^{\dagger} \right)^{k_M}
\ket{0}
\end{displaymath}
and a basis transformation is obtained from the canonical transformation
\begin{displaymath}
b_j^{\dagger}(U) = \sum_{r=1}^M U_{j r} a_r^{\dagger}
%\hspace{2cm} U U^{\dagger} = U^{\dagger} U = \mathbb{I}.
\end{displaymath}
%RMN defined by the unitary $U$ of size $M \times M$.
defined by the  $M \times M$ unitary matrix $U$.
The whole variational ansatz then depends on two sets of parameters: the tensors $\{A_1,\ldots,A_M\}$ and the unitary~$U$. The goal of the algorithm is to find the minimium of the energy with respect to these two sets of parameters, i.e., to calculate
\begin{displaymath}
E = \min_{A_1 \ldots A_M U} \bra{\Psi(A_1,\ldots,A_M,U)} H \ket{\Psi(A_1,\ldots,A_M,U)}.
\end{displaymath}
The idea is to perform the optimizations over the parameter sets $\{A_1,\ldots,A_M\}$ and $U$ consecutively and repeatedly until convergence is reached.

In the following two sections, we describe the two optimization procedures in more detail.

%\begin{displaymath}
%\ket{\phi_{k_1,\ldots,k_M}(U)} =
%\left( b_1^{\dagger}(U) \right)^{k_1}
%\cdots
%\left( b_M^{\dagger}(U) \right)^{k_M}
%\ket{0}
%\end{displaymath}

%\begin{displaymath}
%\ket{\Psi(C,U)} = \sum_{k_1,...,k_M=1}^d C_{k_1 \ldots k_M} \ket{\phi_{k_1,...,k_M}(U)}
%\end{displaymath}

% ====

\subsection{Network Optimization}

\begin{figure}[t]
    \begin{center}
        \includegraphics[width=0.44\textwidth]{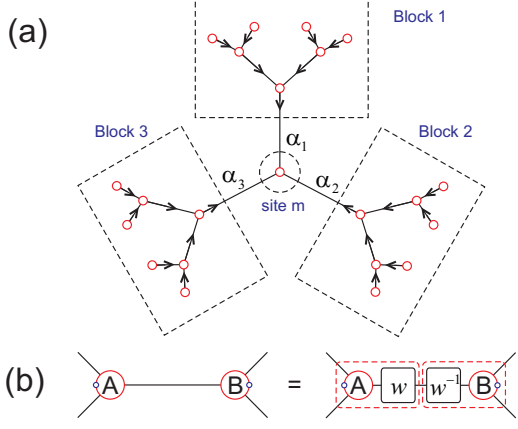}
    \end{center}
    \caption{
       (Color online) (a) Separation of the state into~$z$ blocks plus
      the site under optimization, as described by
      Eq.~(\ref{eqn:blockwavefct}). (b) Natural freedom in the tensor
      network: insertion of a matrix $w$ and $w^{-1}$ at one bond
      leaves the state invariant. The contraction of $A$ with $w$
      forms the new tensor~$A'$ on the left hand side; the contraction
      of $B$ with $w^{-1}$ forms the new tensor~$B'$ at the right hand
      side. 
        }
    \label{fig:orthonormalization}
\end{figure}

First, let us sketch how to optimize the tensors $\{A_1,\ldots,A_M\}$ while keeping the basis fixed.
The minimization of the energy with the constraint that the norm remains constant is
equivalent to minimizing the functional
\begin{displaymath}
F = \bra{\Psi} H \ket{\Psi} - E \left( \scal{\Psi}{\Psi} - 1 \right).
\end{displaymath}
This functional is non-convex with respect to all parameters
$\{A_1,\ldots,A_M\}$. However, due to the tensor network structure of
the ansatz, it is quadratic in the parameters $A_m$ associated with
one lattice site~$m$. Because of this, the optimal parameters $A_m$
can simply be found by solving a generalized eigenvalue problem
$\mathcal{H}_m \vec{A}_m = E \mathcal{N}_m \vec{A}_m$. For a bipartite
network, it is always possible to assume a gauge condition so that
$\mathcal{N}_m = \mathbb{I}$, and thus reduce the generalized
%RMN
eigenvalue problem 
to an ordinary one.\cite{shi06}. %RMN eigenvalue problem~\cite{shi06}. 
We will discuss this in more
detail later in this section. The concept of the algorithm is to do
this one-site optimization site-by-site until convergence is reached. 

The challenge that remains is to calculate the effective Hamiltonian
$\mathcal{H}_m$ of the eigenvalue problem. In principle, this is done
by contracting all indices in the expression for the expectation value
$\bra{\Psi} H \ket{\Psi}$ except those that connect %RMN connecting 
to $A_m$. By
interpreting the tensor $A_m$ as a $d D^z$-dimensional vector
$\vec{A}_m$,
this expression can be written as 
\begin{equation} \label{eqn:Hm}
\bra{\Psi} H \ket{\Psi} = \vec{A}_m^{\dagger} \mathcal{H}_m \vec{A}_m.
\end{equation}
Since
\begin{equation} \label{eqn:Nm}
\scal{\Psi}{\Psi} = \vec{A}_m^{\dagger} \mathcal{N}_m \vec{A}_m
\end{equation}
and $\mathcal{N}_m = \mathbb{I}$, the functional~$F$ attains its
minimum when %RMN as
\begin{displaymath}
\mathcal{H}_m \vec{A}_m = E \vec{A}_m.
\end{displaymath}
Due to the bipartite structure of the tensor network, the calculation
of $\mathcal{H}_m$ can be performed efficiently, i.e., on %RMN in 
a time that scales polynomially with~$M$ and~$D$. 
Assuming that the %RMN a
coordination number~$z$ %RMN throughout,
is the same everywhere,
the scaling will be $M d
D^{z+1}$.

\begin{figure}[t]
    \begin{center}
        \includegraphics[width=0.44\textwidth]{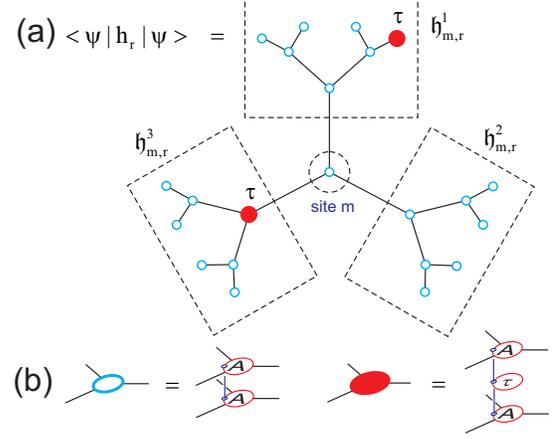}
    \end{center}
    \caption{
       (Color online) (a) Formation of the effective Hamiltonian
      $\mathfrak{h}_{m,r} = \mathfrak{h}_{m,r}^1 \otimes
      \mathfrak{h}_{m,r}^2 \otimes \mathfrak{h}_{m,r}^3$ with respect
      to the interaction $h_r = \tau^{(7)} \otimes \tau^{(15)}$. 
      The
      sites on which the interaction has support are marked in
      red. Each open (filled) circle in the tensor network corresponds to
      the contraction of the layered structure of tensors shown
      in~(b).
        }
    \label{fig:heff}
\end{figure}

% -- orthonormalization --
This procedure is similar to %RMN resembles 
a %RMN an ordinary 
DMRG calculation with~$z$ blocks instead of two, %RMN 2.
where a %RMN Each 
block consists of all of the  sites within one of the branches
emerging from site~$m$ (see Fig.~\ref{fig:orthonormalization}(a)). 
The wave function is then formed as
\begin{equation} \label{eqn:blockwavefct}
\ket{\Psi} =
\sum_{\alpha_1,\ldots,\alpha_z=1}^D
\ket{\varphi_{\alpha_1 \ldots \alpha_z}} \otimes
\ket{\phi_{\alpha_1}^1} \otimes \cdots \otimes \ket{\phi_{\alpha_z}^z}
\, ,
\end{equation}
where %RMN with 
$\ket{\phi_{\alpha}^i}$ ($\alpha=1,\ldots,D$) is %RMN being 
the basis in environment block~$i$ ($i=1,\ldots,z$) and
$\ket{\varphi_{\alpha_1 \ldots \alpha_z}}$ is the state of
site~$m$. 
Since $\mathcal{N}_m$ is obtained from the norm $\scal{\Psi}{\Psi}$ by
contracting all tensors except $A_m$ [see Eq.~(\ref{eqn:Nm})], it
factorizes into a tensor product of~$z$ matrices~$\mathcal{N}_m^i$,
\begin{displaymath}
\mathcal{N}_m = \mathcal{N}_m^1 \otimes \cdots \otimes \mathcal{N}_m^z,
\end{displaymath}
where %RMN with 
each matrix $\mathcal{N}_m^i$ is formed by taking the overlap of the basis states in environment block~$i$:
\begin{displaymath}
\left[ \mathcal{N}_m^i \right]_{\alpha \beta} =
\scal{\phi_{\alpha}^i}{\phi_{\beta}^i}
\end{displaymath}
Obviously, in order to guarantee that $\mathcal{N}_m = \mathbb{I}$,
%RMN it is necessary that 
the basis in each environment block must be orthonormal. This is a
similar requirement as in the DMRG. 
In the tree tensor network, this may be achieved by assuming an
appropriate gauge condition. 
The purpose of this gauge condition is to fix %fixes 
the natural freedom in the tensor network that a matrix
and its inverse can be inserted at any bond, with the matrix being
contracted with the 
first attached tensor and the inverse being contracted with the second
attached tensor, %RMN thereby 
leaving the network invariant (see Fig.~\ref{fig:orthonormalization}(b)). The gauge condition for all sites~$n$ ($n \neq m$) that assures that $\mathcal{N}_m = \mathbb{I}$ is
\begin{equation} \label{eqn:gauge}
\sum_{k \beta_2 \ldots \beta_z}
\left[ A_n^{*} \right]_{\alpha' \beta_2 \ldots \beta_z}^k
\left[ A_n \right]_{\alpha \beta_2 \ldots \beta_z}^k = \delta_{\alpha \alpha'}.
\end{equation}
Here we take the index $\alpha$ to be outgoing, i.e., the branch
attached to this index contains site~$m$, and the %RMN that 
indices $\beta_2 \ldots \beta_z$ to be incoming, i.e., the attached branches 
\emph{do not} contain site~$m$ (see Fig.~\ref{fig:orthonormalization}(a)). 

A stable way to apply this gauge condition is to ``orthonormalize''
the tensors from %RMN the 
outside to inside. That is, starting with the
tensors on the leaves of the tree, we take into account
condition~(\ref{eqn:gauge}) 
by QR-decomposing $\left[ A_n \right]_{\alpha}^k$, i.e., by %RMN
writing it as
\begin{displaymath}
\left[ A_n \right]_{\alpha}^k = \sum_{\alpha'} \left[ Q_n \right]_{\alpha'}^k R_{\alpha' \alpha}.
\end{displaymath}
The unitary matrix $\left[ Q_n \right]_{\alpha'}^k$ is the new ``orthonormalized''
tensor at site~$n$. In order to keep the tensor network invariant,
$R_{\alpha' \alpha}$ must be contracted with the
tensor on the first inner shell that is connected to the tensor at site~$n$.
%RMN Iteratively, 
This procedure continues iteratively  with the tensors on the first
inner shell, the second inner shell, and so on until site~$m$ is
reached.

\begin{figure}[t]
    \begin{center}
        \includegraphics[width=0.44\textwidth]{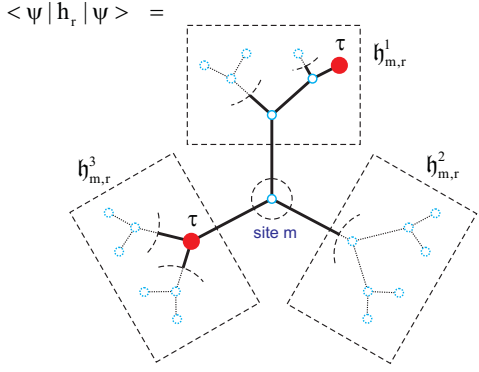}
    \end{center}
    \caption{
      (Color online) Branches with no support (marked by dotted lines
      and circles), which %RMN
      yield the identity when contracted and therefore do not have to be taken
      into account in calculating %RMNthe calculation of 
      the effective Hamiltonian. 
    }
    \label{fig:cropping}
\end{figure}

\begin{figure}[t]
    \begin{center}
        \includegraphics[width=0.44\textwidth]{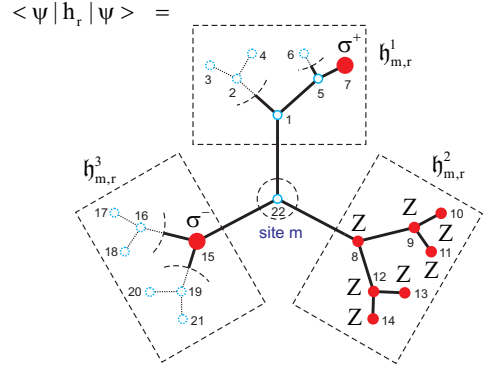}
    \end{center}
    \caption{
       (Color online) Formation of the effective Hamiltonian $\mathfrak{h}_{m,r} = \mathfrak{h}_{m,r}^1 \otimes \mathfrak{h}_{m,r}^2 \otimes \mathfrak{h}_{m,r}^3$ with respect to the fermionic interaction $h_r = a_7^{\dagger} a_{15}$. The sites on which the interaction has support are marked in red.
        }
    \label{fig:hefffermions}
\end{figure}

% -- effective hamiltonian --
Thus, by assuring that the gauge condition is always satisfied in the
course of the algorithm, the only term that must %RMN has to 
be calculated is
the effective Hamiltonian~$\mathcal{H}_m$. This term is obtained, as
can be gathered from Eq.~(\ref{eqn:Hm}), by contracting all tensors
except $A_m$ in the expectation value~$\bra{\Psi} H \ket{\Psi}$. 
Since %RMN As 
the Hamiltonian is a sum of interaction terms %RMN,
\begin{displaymath}
H = \sum_{r=1}^R h_r,
\end{displaymath}
with $h_r$ being a tensor product of matrices, e.g., $h_r = \tau^{(7)} \otimes \tau^{(15)}$ for a two-body interaction acting on sites~$7$ and $15$, the effective Hamiltonian splits into a sum of effective Hamiltonians~$\mathfrak{h}_{m,r}$ with respect to a single interaction term~$h_r$:
\begin{displaymath}
\bra{\Psi} h_r \ket{\Psi} = \vec{A}_m^{\dagger} \mathfrak{h}_{m,r}
\vec{A}_m \, .
\end{displaymath}
Due to %RMN Because of 
the structure~(\ref{eqn:blockwavefct}) of the TTNS, each
effective Hamiltonian~$\mathfrak{h}_{m,r}$ factorizes into a tensor product of~$z$ matrices
\begin{displaymath}
\mathfrak{h}_{m,r} = \mathfrak{h}_{m,r}^1 \otimes \cdots \otimes
\mathfrak{h}_{m,r}^z \, ,
\end{displaymath}
where %RMN with 
each matrix $\mathfrak{h}_{m,r}^i$ corresponds to the matrix elements of $h_r$ with respect to the basis in environment block~$i$:
\begin{displaymath}
\left[ \mathfrak{h}_{m,r}^i \right]_{\alpha \beta} =
\bra{\phi_{\alpha}^i} h_r \ket{\phi_{\beta}^i} \, .
\end{displaymath}
Graphically, the evaluation of $\bra{\phi_{\alpha}^i} h_r
\ket{\phi_{\beta}^i}$ corresponds to the contraction of a
three-layered %RMN $3$~layered
tensor network according to the structure of the branch in block~$i$,
as depicted in Fig.~\ref{fig:heff}. This network can be contracted
efficiently by starting from the leaves and working in the inward direction. The numerical effort for contracting one additional site into the network scales as $dD^{z+1}$, so that the total effort scales as $dD^{z+1}$ times the number of sites in the block.

Of course, if $h_r$ has no support in environment block~$i$,
$\mathfrak{h}_{m,r}^i$ is equal to the identity because 
%RMN of the orthonormal choice of the basis 
the basis is chosen to be orthogonal in each environment block. Because of
this, the calculation simplifies significantly. For example, each
two-site interaction has support in at most two blocks. This means
that at most two effective Hamiltonians~$\mathfrak{h}_{m,r}^i$ have to
be calculated (for each interaction term); all others are equal to the
identity. 
Since the orthonormalization of the state is applied iteratively from
the leaves inwards to the optimized site~$m$, the calculation of
$\mathfrak{h}_{m,r}^i$ simplifies substantially, %RMN a lot, 
as well. Within each block,
the contraction of all subbranches on which $\mathfrak{h}_{m,r}^i$ has
no support %RMN is 
automatically yields %RMN equal to 
the identity. Thus, for the
example of a two-site interaction, it is sufficient to take into
account the sites on the path connecting the two sites (see
Fig.~\ref{fig:cropping}). The treatment of long-range interactions on
a tree is therefore not significantly %RMN a lot 
more complicated than the treatment of
long-range interaction on a chain in the DMRG. 

% -- symmetries --
At first glance, %RMN sight, it seems 
it might seem that the advantage of ``cropping'' all
subbranches with no support is lost when switching to fermions. This
is because fermionic interactions with local support are turned into
interactions with nonlocal support in the spin picture via the
Jordan-Wigner transformation. The two-particle fermionic interaction
$a_7^{\dagger} a_{15}$, for example, is turned into the spin interaction
\begin{displaymath}
a_7^{\dagger} a_{15} = \sigma_{+}^{(7)} \left[ \prod_{7<n<15} Z^{(n)} \right] \sigma_{-}^{(15)}
\end{displaymath}
that has support on 9 sites,
where %RMN with 
$Z=-\sigma_z$ and $\sigma_{+}$, $\sigma_{-}$, $\sigma_z$ denote the
Pauli operators (see Fig.~\ref{fig:hefffermions}). 
However, as we will show, local fermionic interactions can be treated
in the tree network with the same effort as local spin interactions by
including the $Z_2$-symmetry in the ansatz. 
This is the fermion number parity
in the fermionic language.

The $Z_2$-symmetry can be incorporated into the ansatz by making the
tensors $\left[ A_m \right]^{k}_{\alpha_1 \ldots \alpha_z}$
block-diagonal.
 That is, each virtual index is split into an index
pair $(\alpha_i, p_i)$, with $p_i$ carrying the parity information. 
By stipulating that $p_1 \oplus \cdots \oplus p_z \oplus k = 1$ (where
%RMN with
$\oplus$ denotes summation modulo~$2$), it is possible to
``move'' the operator $Z$ that acts on the physical index of $A_m$ to
the virtual parity indices~$p_i$: 
\begin{equation} \label{eqn:parityrelation}
Z_{k \tilde{k}} \left[ A_m \right]^{\tilde{k}}_{ \alpha_1 p_1 \cdots \alpha_z p_z } =
\left[ A_m \right]^{k}_{\alpha_1 \tilde{p}_1 \cdots \alpha_z
  \tilde{p}_z} Z_{\tilde{p}_1 p_1} \cdots Z_{\tilde{p}_z p_z} \, .
\end{equation}
From this relation, it immediatly follows that
\begin{equation} \label{eqn:Zpsi}
Z \otimes \ldots \otimes Z \ket{\Psi} = \ket{\Psi}
\end{equation}
and thus that the state is $Z_2$-symmetric. This is because the
left-hand side of Eq.~(\ref{eqn:Zpsi}) corresponds to the contraction
of~$Z$'s to the physical indices of all tensors~$A_m$. After moving
all~$Z$'s to the virtual bonds, %RMN on each bond 
the operator~$Z$ appears
twice on each bond .  
Thus, since $Z^2 = \mathbb{I}$, the state $\ket{\Psi}$ remains unchanged.

\begin{figure}[t]
    \begin{center}
        \includegraphics[width=0.44\textwidth]{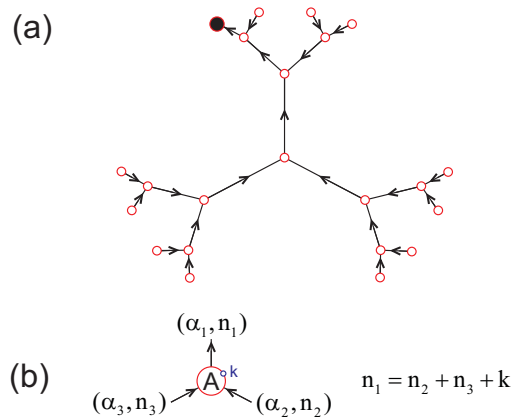}
    \end{center}
    \caption{
       (Color online) (a) Possible choice of an ordered tree graph with each site
      having two incoming and one outgoing bond. 
      The sink site is marked in black.
      The tensors~$A_m$ associated with each site with
      virtual indices consisting of pairs~$(\alpha_i,n_i)$ have the
      structure %RMN as 
      shown in~(b).
        }
    \label{fig:particlesymmetry}
\end{figure}

Using the same idea, we can immediately enforce a more restrictive
symmetry that is fulfilled by all fermionic Hamiltonians:
the charge symmetry $U(1)$, i.e. the conservation of the number of particles.
For this, the tree graph has to be made
directed (see Fig.~\ref{fig:particlesymmetry}), so that all sites
(except one) have~$z-1$ incoming and one outgoing bond. The exception
is the sink site with~$z$ incoming bonds. Thus, each virtual index of
a tensor~$A_m$ is equipped with the additional information of whether
it is ``incoming'' or ``outgoing''. We assume that the index~$1$ is
always the outgoing index in the following. As before, each virtual
index is split into an index pair $(\alpha_i,n_i)$. In addition, we
require that $n_1 = n_2 + \ldots + n_z + k$. Thus, for $i=2,\ldots,z$,
$n_i$ counts the number of particles within the branch attached to
index~$i$.  The index $n_1$, on the other hand, is equal to the number
of particles in all the branches plus the number of particles at
site~$m$. Since the parity can be immediately derived from the
particle-number information, a similar relation
as Eq.~(\ref{eqn:parityrelation}) holds, namely
\begin{equation} \label{eqn:parityrelation2}
Z_{k \tilde{k}} \left[ A_m \right]^{\tilde{k}}_{ \alpha_1 n_1 \cdots \alpha_z n_z } =
\left[ A_m \right]^{k}_{\alpha_1 \tilde{n}_1 \cdots \alpha_z \tilde{n}_z} \tilde{Z}_{\tilde{n}_1 n_1} \cdots \tilde{Z}_{\tilde{n}_z n_z}
\end{equation}
with $\tilde{Z}_{\tilde{n} n} = \delta_{\tilde{n} n} (-1)^n$. Thus, as
before, $Z$ acting on the physical index can be moved to the virtual
particle-number bonds and, since $\tilde{Z}^2 = \mathbb{I}$, the state
is $Z_2$-symmetric.

\begin{figure}[t]
    \begin{center}
        \includegraphics[width=0.44\textwidth]{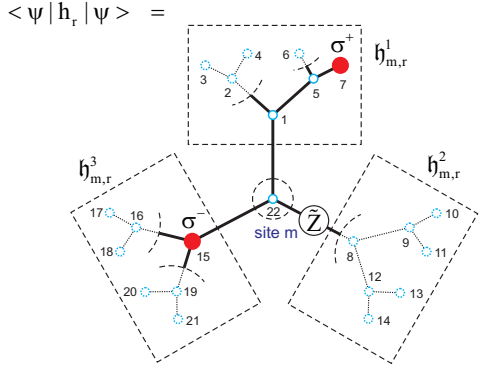}
    \end{center}
    \caption{
       (Color online) Formation of the effective Hamiltonian $\mathfrak{h}_{m,r} =
        \mathfrak{h}_{m,r}^1 \otimes \mathfrak{h}_{m,r}^2 \otimes
        \mathfrak{h}_{m,r}^3$ with respect to the fermionic
        interaction $h_r = a_7^{\dagger} a_{15}$ with parity symmetry taken into account. The sites on which the interaction has support are marked in red. All branches marked by dotted lines and circles yield the identity when contracted. The parity operator~$\tilde{Z}$ is contracted to the virtual bond.
        }
    \label{fig:hefffermionswithparity}
\end{figure}

The advantage of taking into account the particle number is twofold:
on the one hand, the block structure of the tensors reduces the
numerical effort considerably: a bond-dimension of $\chi = N D$ can be
treated with an effort of order $N^z D^{z+1}$ instead of $\chi^{z+1}$
for %RMN in case of 
a non-symmetric ansatz. 
Here $\chi$ is the full bond dimension, so it is equivalent to the number of block states
kept in a DMRG procedure.
On the other hand, the calculation of the effective Hamiltonians
$\mathfrak{h}_{m,r}$ with respect to an interaction with support on a
few sites only, e.g., a two-particle or four-particle interaction, is
simplified. The main idea is depicted in %RMN can be gathered from
Fig.~\ref{fig:hefffermionswithparity} for the
interaction~$a_7^{\dagger} a_{15}$: with an appropriately chosen
numbering of the fermions, each subbranch that has no fermionic
support either has only %RMN all 
identities acting on the sites or only %RMN all
operators~$Z$. The subbranches including only identities simplify to
the identity because of the orthonormalization of the state. The
$Z$ operators can be moved, using
relation~(\ref{eqn:parityrelation2}), to the virtual bonds, and all of
them except one cancel (see Figs.~\ref{fig:hefffermions}
and~\ref{fig:hefffermionswithparity}). 
What remains is a subbranch
that includes %RMN including 
only identities, which reduces to %RMN equals 
the identity because of the
orthonormalization of the state. Thus, for a fermionic two-site
interaction, it is sufficient to take into account the path connecting
the two sites. 
In this way, %RMNBy this, 
the treatment of long-range fermionic
interactions is feasible with the same numerical effort as the
treatment of long-range spin interactions.

% -- time evolution --
Using the same scheme as for minimizing the energy, the efficient
simulation of time evolution is also possible. For this, the time
evolution is split up into small steps of duration $\delta t$. For
$\delta t \ll 1$ and a starting state $\ket{\Psi_0}$ of the form of a
TTNS, the time-evolved state $e^{-i H \delta t} \ket{\Psi_0}$ is a
TTNS as well.  
However, the virtual dimension is multiplied by a factor
$\kappa>1$. 
In order to prevent %RMN that 
the bond dimension from increasing %RMN increases
exponentially with time, the TTNS has to be approximated by a TTNS
with a reduced virtual dimension $\ket{\Psi}$ after every time step,
i.e., the functional
\begin{displaymath}
K = \lVert e^{-i H t} \ket{\Psi_0} - \ket{\Psi} \rVert^2
\end{displaymath}
must %RMN has to 
be minimized. The optimization with respect to a single site~$m$ leads to the system of linear equations
$\mathcal{N}_m \vec{A}_m = \vec{w}_m$ (where $\mathcal{N}_m$ can again
be set equal to identity by using the appropriate gauge)
with
\begin{displaymath}
\bra{\Psi} e^{-i H t} \ket{\Psi_0} = \vec{A}_m^{\dagger} \vec{w}_m
\end{displaymath}
and thus can be performed efficiently with the scaling $M D^{z+1}$, as
before. 
Clearly, %RMN Of course, 
all previous considerations regarding %RMN about 
particle-number
symmetry can also be adapted to the case of time evolution.

% ====

\subsection{Orbital Optimization}

As mentioned previously, the entanglement properties of the system
depend critically on the choice of the basis.  
Our goal is to find a basis in which entanglement is localized as much
as possible at the sites of the tree network. 
Such a choice of basis guarantees that a given %RMN same 
precision 
can be attained %RMN
with a smaller
virtual dimension~$D$, and thus with less %RMN a smaller 
computational effort. 
In QC-DMRG the optimization of the basis is fundamental and
has been %RMN 
used in much %RMN many 
previous work.~\cite{white02,zgid08,zgid08a,ghosh08,luo10}  

In contrast to %RMN the 
previous work, our approach aims to find %RMN at finding 
the optimal basis that
can be obtained by a canonical transformation of %RMN between 
the fermionic modes. 
The canonical transformation is defined by a  $M \times M$ 
unitary matrix $U$. Since the number of parameters is relatively %RMN very 
small, a gradient
search applied to %RMN on 
the expectation value of the energy, 
\begin{displaymath}
E(U) = \bra{\Psi(U)} H \ket{\Psi(U)},
\end{displaymath}
is certainly feasible. 
Since %RMN Of course, 
$E(U)$ is a non-convex function of the parameters~$U$, %RMN and
it is a highly non-trivial problem to find the absolute minimum. The
idea is to perform the gradient search multiple times in the course of
the algorithm, e.g., after each optimization sweep of the tensor
network, and improve the energy at each gradient search by only a
small amount. %RMN quantity. 
This will eventually adapt the orbitals optimally to
the tree tensor network. 

There are two ways to implement the basis transformation: one based on
%RMN on the basis of 
the state and the other based on %RMN or on the basis of 
the Hamiltonian. We have applied the
basis transformation to the Hamiltonian. 
For %RMN In the case of 
the fermionic Hamiltonian with long-range interactions,
\begin{displaymath}
H = \sum_{i j} T_{i j} a_{i}^{\dagger} a_{j} +
\sum_{i j k l} V_{i j k l} a_{i}^{\dagger} a_{j}^{\dagger} a_{k} a_{l}
\, ,
\end{displaymath}
that appears, e.g., in quantum chemistry, in momentum-space
descriptions of the Hubbard model, or in descriptions of the Hubbard model
on higher dimensional lattices, the function $E(U)$ can be be
expressed as 
\begin{displaymath}
E(U) =
\sum_{i j} \tilde{T}(U)_{i j} \expect{a_i^{\dagger} a_j} +
\sum_{i j k l} \tilde{V}(U)_{i j k l} \expect{a_i^{\dagger} a_j^{\dagger} a_k a_l}
\end{displaymath}
with
\begin{eqnarray*}
\tilde{T}(U) & = & U T U^{\dagger} \\
\tilde{V}(U) & = & (U \otimes U) V (U \otimes U)^{\dagger}.
\end{eqnarray*}
The correlation functions $\expect{a_i^{\dagger} a_j}$ and
$\expect{a_i^{\dagger} a_j^{\dagger} a_k a_l}$ are calculated with
respect to the original state and are not dependent on the
parameters in $U$. %RMN
With the function $E(U)$ in this form, its gradient
can be calculated explicitly. Both quantities can be evaluated
efficiently for different parameter sets~$U$, which makes %RMN will make 
the gradient search feasible.

The gradient search that is used for the orbital optimization 
typically finds %RMN has the property to find 
the local minima in the vicinity of the starting
point of the optimization. 
In order to avoid falling into %RMN to end up in 
the same minimum, we shift the starting point in a random direction by 
an appropriately chosen amount. %RMN a certain magnitude. 
This has the consequence that the gradient search
generally falls into %RMN ends up in 
another local minimum that may have a higher
energy. 
However, it assures that the orbitals change considerably and
that, since the orbital optimization is performed repeatedly 
interspersed with %RMN with intermediate 
network optimization, the energy decreases notably in the
course of the algorithm.

% =============================================

\section{Numerical Results} \label{sec:results}

%RMN The algorithm consisting of two optimization procedures described
%previously -- the optimization of the tensor network and the
%optimization of the basis according to this network -- we have applied
%to several models. The results we discuss in the following. 

We have applied the algorithm consisting of two optimization
procedures described above, 
the optimization of the tensor network and the
optimization of the basis according to this network, 
to several models. 
In this section, we discuss the results.

% ====

\subsection{2-d Heisenberg model}

\begin{figure}
    \begin{center}
        \includegraphics[width=0.5\textwidth]{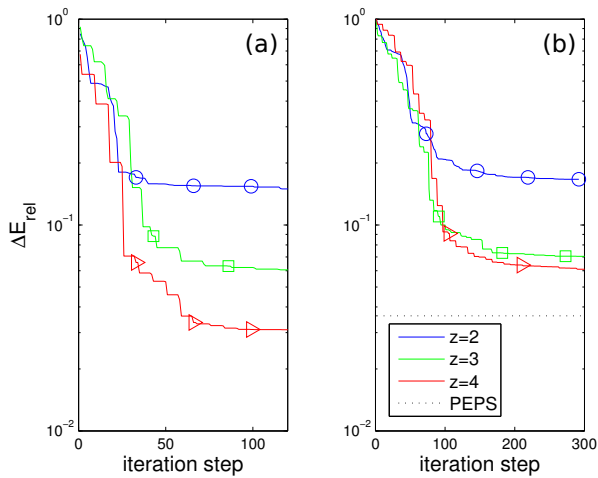}
    \end{center}
    \caption{
       (Color online) Relative error in the energy for Heisenberg model on a $4 \times 4$-lattice (a) and 
	a $6 \times 6$-lattice (b) as a function of the optimization steps.
	The calculations were performed with a fixed virtual dimension of $D=4$ and tree tensor networks with
	coordination numbers $z=2$ (blue curve), $z=3$ (green curve) and $z=4$ (red curve).
        }
    \label{fig:heisenberg}
\end{figure}

First, let us consider the tree tensor network optimization only and
show using %RMN at hands of 
the 2-d Heisenberg model how results improve with a TTNS ansatz compared to a one-dimensional MPS ansatz. 
The relative error in the energy of a system consisting of $4 \times 4$ spins as a function of the optimization step is shown in the left hand side of Fig.~\ref{fig:heisenberg} for the MPS ansatz and the TTNS ansatz with $z=3$ and $z=4$. In all calculations, a fixed virtual dimension of~$D=4$ is used. We have assigned a spin to each node in the tree network in such a way that two arbitrary spins are connected by the smallest possible number of bonds. As can be seen, the precision increases considerably with increasing coordination number.

%RMN A similar statement is true 
For larger systems such as for $6 \times 6$ spins, shown in the
right hand side of Fig.~\ref{fig:heisenberg}, the energy is
plotted as a function of the optimization steps for $z=2$, $z=3$, and
$z=4$. The virtual dimension is fixed to $D=4$. As before, the energy
decreases considerably and approaches the PEPS result for $D=4$,
indicated by the dotted line which, however, is obtained with a much
larger %RMN bigger 
numerical effort.\cite{murgverstraete07}
% ====

\subsection{2-d interacting spinless fermions}

\begin{figure}
    \begin{center}
        \includegraphics[width=0.5\textwidth]{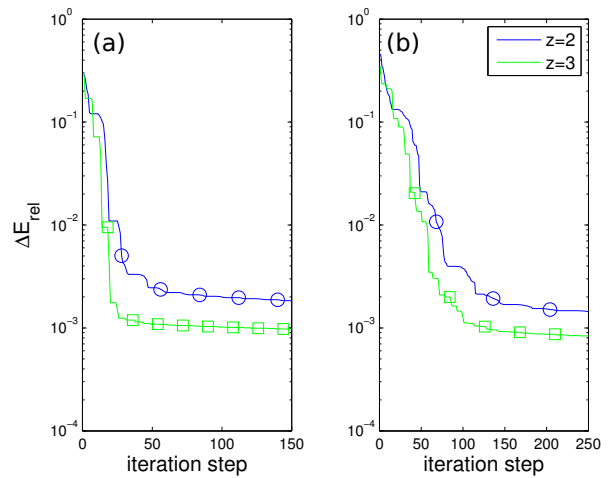}
    \end{center}
    \caption{
       (Color online) Relative error in the energy for the interacting spinless 
	fermion model on (a) a $4 \times 4$-lattice and (b) a $6 \times 6$-lattice as a function of the optimization steps.
	The calculations were performed with a fixed virtual dimension of $D=4$ and tree tensor networks with
	coordination numbers $z=2$ (blue curve) and $z=3$ (green curve).
	The chosen parameters are $J=1$, $U=0.5$ and the number of fermions is fixed to $N=3$.
        }
    \label{fig:spinlessfermihubbard2d}
\end{figure}

The TTNS Ansatz also perfoms well on two-dimensional fermionic models. We have applied the TTNS algorithm to a system of interacting fermions on a two-dimensional lattice, described by the Hamiltonian
\begin{displaymath}
H = -J \sum_{<i,j>} a_i^{\dagger} a_{j} +
U \sum_{<i,j>} \hat{n}_i \hat{n}_{j}
\end{displaymath}
with $\hat{n}_i = a_i^{\dagger} a_i$. The boundary conditions are assumed to be periodic. As can be gathered from Fig.~\ref{fig:spinlessfermihubbard2d}, the ground-state energy improves with increasing coordination number~$z$. The figure shows the relative error in the energy as a function of the iteration step for $z=2$ and $z=3$ for a $4 \times 4$-lattice in the left part, and the ground state energy as a function of the iteration step for a $6 \times 6$-lattice in the right part. For the calculations, we have used the virtual dimension~$D=4$. We have chosen the parameters $J=1$ and $U=0.5$ and have fixed the number of fermions to $N=3$.

Thus, a TTNS Ansatz might be useful for the study of higher
dimensional models of small size because the effective long-range
interactions are represented better in a tree than in a chain and the
numerical effort is relatively low compared to a PEPS calculation.

% ====

\subsection{1-d interacting spinless fermions}

\begin{figure}
    \begin{center}
        \includegraphics[width=0.5\textwidth]{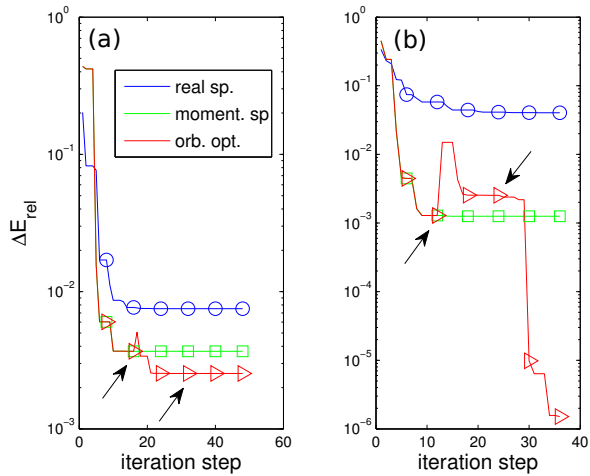}
    \end{center}
    \caption{
      (Color online) Relative error in the energy for interacting
      spinless fermions  
      with~$M=7$ sites as a function of the
      optimization steps for the TTNS ansatz with $z=2$ (a) and $z=3$ (b).
      Here $U=1$, the particle
      number is fixed to $N=3$, 
      and the virtual dimension is assumed to be $D=2$.
      Results with orbital optimization %(red line) 
      are compared
      to the results obtained in the real-space
      basis %(blue line) 
      and in the momentum-space basis. % (green line).
      The steps at which orbital optimizations are performed are marked with arrows.
    }
    \label{fig:spinlessfermihubbard}
\end{figure}

In order to assess the effectiveness of the orbital optimization, we have
applied the algorithm to a simple fermionic model, the one-dimensional
interacting spinless fermion model
\begin{displaymath}
H = -J \sum_{i=1}^{M} \left( a_i^{\dagger} a_{i+1} + \textrm{h.c.} \right) +
U \sum_{i=1}^{M} \hat{n}_i \hat{n}_{i+1} \, ,
\end{displaymath}
where $\hat{n}_i = a_i^{\dagger} a_i$. We assume periodic boundary conditions, i.e., $a_{M+1}=a_1$.
This model can be mapped to the XXZ spin chain via a
Jordan-Wigner transformation.
In this model, it is known that the choice of the basis has a big
effect on the precision of the DMRG or TTNS calculation. 
For $U \to \infty$, the ground state is a product state in the position
representation, and thus optimally represented by a tensor network
with $D=1$ in this basis. 
For $U=0$, on the other hand, the ground
state can be represented as a direct product in momentum space.
Thus, the momentum-space basis is clearly best suited in this limit. 
For intermediate $U$, one might expect a basis intermediate between real
and momentum space to represent the entanglement properties optimally;
our goal is to find such a basis automatically by carrying out orbital
optimization.

The results of calculations incorporating the optimization procedure
are displayed in Fig.~\ref{fig:spinlessfermihubbard} for $z=2$ and $z=3$, both
on systems of $M=7$ sites and $N=3$ particles with a fixed virtual
dimension of~$D=2$ and parameters~$U=1$. 
%In addition, we have fixed the number of
%particles to~$N=3$. 
As before, the relative error in the energy as a
function of the optimization step is plotted. 
For comparison, calculations performed in the position representation
and in the momentum representation are also shown. 
As can be seen, the energy improves significantly in the
course of the optimization; the improvement is
three orders of magnitude for the $z=3$ case.

\begin{figure}[t]
    \begin{center}
        \includegraphics[width=0.5\textwidth]{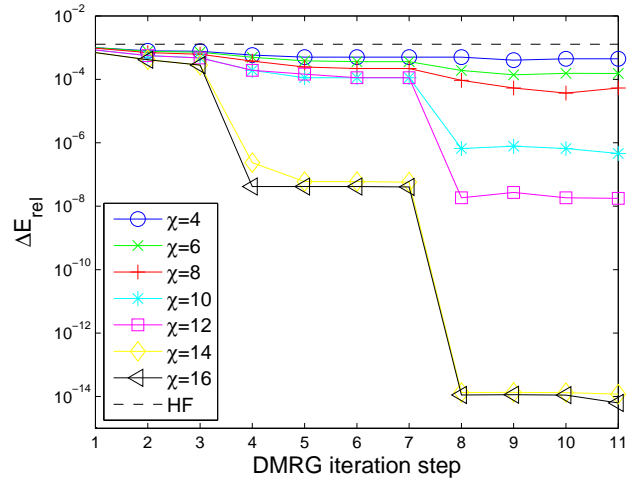}
    \end{center}
    \caption{
    (Color online) The relative error in the ground-state energy for the
        Beryllium atom as a function of DMRG iteration steps for
        various values of the number of DMRG block states.
        The dashed line corresponds to the Hartree-Fock energy.
        }
    \label{fig:be48_dmrg}
\end{figure}

% ====

\subsection{Quantum Chemical systems}

In this section, we compare numerical results for quantum
chemical systems obtained using the QC-DMRG and TTNS
methods. 
In these applications, the electron-electron correlation
is taken into account by an iterative
procedure that minimizes the Rayleigh quotient corresponding
to the Hamiltonian describing the electronic
structure of the molecule, given by
\begin{equation} \label{eqn:Hquantchem}
H = \sum_{i j \sigma} T_{i j} a_{i \sigma}^{\dagger} a_{j \sigma} +
\sum_{i j k l \sigma \sigma'} V_{i j k l} a_{i \sigma}^{\dagger} a_{j \sigma'}^{\dagger} a_{k \sigma'} a_
{l \sigma}
\end{equation}
and thus determines the full-CI wave function. 
In Eq.~(\ref{eqn:Hquantchem}),
$T_{i j}$ denotes the matrix elements of the one-particle
Hamiltonian, which is comprised of the kinetic energy and the external
electric field of the nuclei, and $V_{i j k l}$ stands for the matrix
elements of the electron repulsion operator, defined as
\begin{displaymath}
V_{i j k l} = \int d^3 x_1 d^3 x_2
\Phi_i^{*} (\vec{x}_1) \Phi_j^{*} (\vec{x}_2)
\frac{1}{\vec{x}_1-\vec{x}_2}
\Phi_k (\vec{x}_2) \Phi_l (\vec{x}_1) \, .
\end{displaymath}
We obtain the Hartree-Fock orbitals in a given basis of Gaussian
orbitals and transform the matrix elements 
$T_{i j}$ and $V_{i j  k l}$ 
to the Hartree-Fock basis using
the MOLPRO program package,\cite{molpro} which we also use to obtain
the full-CI energies used as a benchmark.

\begin{figure}[t]
    \begin{center}
        \includegraphics[width=0.5\textwidth]{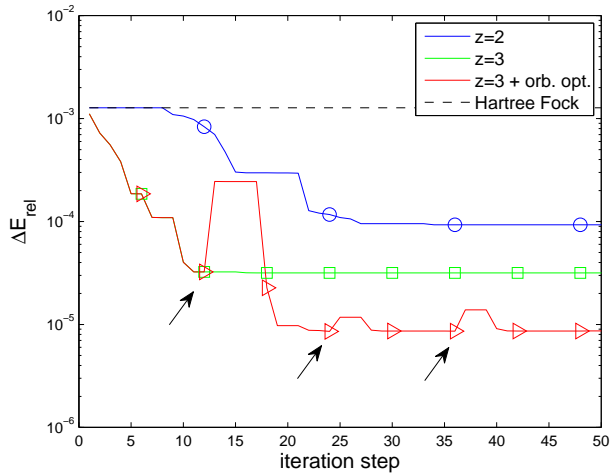}
    \end{center}
    \caption{
      (Color online) Relative error in the ground-state energy for the
      Be atom as 
      a function of the optimization step for $D=2$.
      The results obtained using a TTNS ansatz with $z=3$ plus orbital
      optimization 
      are compared to the TTNS-results
      without orbital optimization 
      For comparison, the results with a MPS ansatz (corresponding to
      $z=2$) are also included. 
      The steps at which orbital optimizations are performed are
      marked with arrows. 
        }
    \label{fig:be84_ttns}
\end{figure}

In the QC-DMRG, a one-dimensional
chain is built up from the atomic or
molecular orbitals obtained from a suitable
mean-field or MCSCF calculation. 
The tree network is constructed similarly, but there is greater
freedom to form the proper structure of the network.
The two-orbital mutual information\cite{rissler06}
provides a good starting configuration.
A general approach to reduce entanglement is to form
the network by placing the highly correlated
orbitals at or near the center of the tree and
less correlated orbitals at the boundary sites.

\begin{figure}[t]
    \begin{center}
        \includegraphics[width=0.5\textwidth]{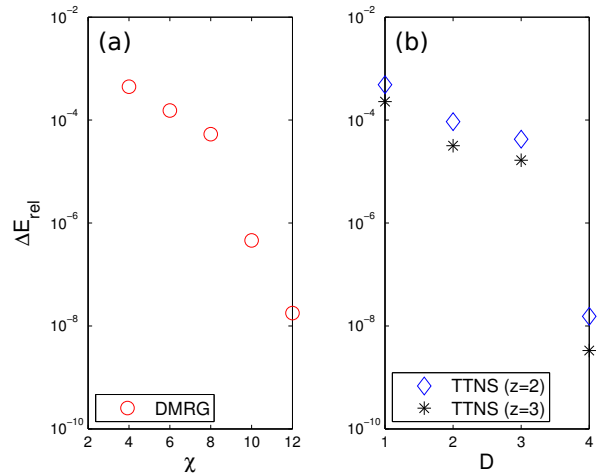}
    \end{center}
    \caption{
    (Color online) Relative error in the ground-state energy for the
        Beryllium atom as a function of the full bond dimension $\chi$
        obtained using (a) the 
	DMRG  and (b) as a function of the virtual dimension~$D$
        obtained from a TTNS with $z=2$ and $z=3$. 
        }
    \label{fig:be48_dmrg_ttns}
\end{figure}

In Fig.~\ref{fig:be48_dmrg}, we plot the relative error in the
ground-state energy for the Beryllium atom as a function of DMRG
iteration steps for various fixed values of the DMRG block states. 
Corresponding data gathered after the fourth DMRG sweep
is shown in Fig.~\ref{fig:be48_dmrg_ttns}.
In this calculation, four
electrons have been placed
on eight orbitals.
The Hartree-Fock energy is~$-14.351880250000$,
while the full-CI energy is~$-14.37016558404629$. 
%
%(The $z = 2$
%limit corresponds to the one-dimensional topology, i.e., to the
%DMRG scheme.)
%}
%\label{tab:be_energy}
%\end{table}
%
%
Figure.~\ref{fig:be84_ttns} depicts the relative
error as a function of the optimization step for the Be atom,
calculated using a TTNS
ansatz with coordination number~$z=3$.
The cases with and without orbital optimization are
considered separately, and the $z=2$ results are included for
comparison. 
As can be seen, the relative error is 
considerably smaller  for $z=3$ than for $z=2$, and there
is a significant gain in precision when orbital optimization is taken
into account. 
In Fig.~\ref{fig:be48_dmrg_ttns}, we display the dependence of the
$z=2$ and $3$ calculations on bond dimension
$D$.
As can be seen, the DMRG and TTNS calculations with
$z=2$ yield similar accuracy, while the
$z=3$ TTNS calculation is significantly more accurate for a given bond
dimension.

It should be noted that there is currently a discrepancy between the
QC-DMRG and the TTNS calculations in the speed of convergence with
optimization step.
Since our QC-DMRG code is
highly optimized, the calculation converges faster than that in the present
version of the TTNS method. 
As can be seen in Fig.~\ref{fig:be48_dmrg}, for example,
the QC-DMRG ground-state energies are always
below the Hartree-Fock energy, and the active space is
extended dynamically using CI-based expansion techniques 
(CI-DEAS)~\cite{legeza03} (DEAS). 
In general, there is no fundamental difficulty in incorporating
optimizations such as the CI-DEAS into the TTNS method.

% =============================================

\section{Conclusion} \label{sec:conclusions}

In this paper, we have 
described and applied a method, 
the tree tensor network state method, to
treat strongly correlated systems with long-range interactions on a
tree network with arbitrary coordination number $z$.
Our approach is based on a  
tensor product state 
ansatz that generalizes a DMRG-like matrix product state  to $z$
rather than two blocks.
The number of virtual bonds required to connect two arbitrary sites scales
logarithmically with the number of sites in TTNS, in contrast to 
the linear scaling of a one-dimensional topology.\cite{shi06}
In this sense, our TTNS method has a lower computational cost than
currently used DMRG-based methods.
We have also incorporated optimization of the single-particle orbitals
in our method, treating the case of a procedure for tranforming the 
basis that smoothly interpolates
between real space and momentum space.
We have tested our method using numerical calculations
on various systems with local and nonlocal interactions, including
the two-dimensional Heisenberg lattice, the momentum-space version of
the 1-d interacting spinless fermion model, and small quantum
chemical systems. 
For the quantem chemical systems, we have compared TTNS results to
those of DMRG calculations.

Since the TTNS approach 
is defined on a bipartite network,  
previous algorithmic
developments and optimizations procedures developed in the context of
the quantum chemistry version of DMRG can also be integrated
into the TTNS method. 
Such optimizations include  
dynamic adjustment of the bond dimension,\cite{legeza03} 
orbital optimization, ~\cite{white02,yanai06,yanai10,zgid08,zgid08a,ghosh08,luo10}
and initialization
procedures based on CI expansions,\cite{legeza03a,moritz06,moritz07} among
others. 
Incorporation of these elements
into the TTNS approach will be carried out in future work. 
In light of the promising features of the new method, we expect it to
provide a viable alternate means of treating atoms and molecules efficiently
in the near future.
%Therefore, the Hartree-Fock
%energy is reached only after a few iteration steps. 
%Implementation of various features
%into the TTNS approach will be part of future projects. These inlcude the
%dynamic selection of block states~\cite{legeza03a,legeza04} (DBSS) used in
%DMRG to a dynamic adjustment of bond dimensions and varying the coordinaton
%number according to the two-site entanglement matrix. 

% =============================================

\acknowledgments{
We thank I.~Cirac, N.~Schuch, and S.~R.~White for very
valuable discussions.
V.~M.\ and F.~V.\ acknowledge support from the SFB projects
FoQuS and ViCoM, the European projects Quevadis, and the ERC grant
Querg.
\"O. L. acknowledges support from
the Alexander von Humboldt foundation and from the 
Hungarian Research Fund(OTKA) through Grant Nos.~K68340 and
K73455.
We would especially like to thank the Erwin-Schr\"odinger-Institut in
Vienna for its hospitality during the Quantum Computation and Quantum
Spin Systems workshop in 2009, where many fruitful discussion took
place.
}

% =============================================

%\bibliography{bibliography}

\begin{thebibliography}{10}%
\makeatletter
\providecommand \@ifxundefined [1]{%
 \ifx #1\undefined \expandafter \@firstoftwo
 \else \expandafter \@secondoftwo
\fi
}%
\providecommand \@ifnum [1]{%
 \ifnum #1\expandafter \@firstoftwo
 \else \expandafter \@secondoftwo
\fi
}%
\providecommand \enquote [1]{``#1''}%
\providecommand \bibnamefont  [1]{#1}%
\providecommand \bibfnamefont [1]{#1}%
\providecommand \citenamefont [1]{#1}%
\providecommand\href[0]{\@sanitize\@href}%
\providecommand\@href[1]{\endgroup\@@startlink{#1}\endgroup\@@href}%
\providecommand\@@href[1]{#1\@@endlink}%
\providecommand \@sanitize [0]{\begingroup\catcode`\&12\catcode`\#12\relax}%
\@ifxundefined \pdfoutput {\@firstoftwo}{%
 \@ifnum{\z@=\pdfoutput}{\@firstoftwo}{\@secondoftwo}%
}{%
 \providecommand\@@startlink[1]{\leavevmode}%
 \providecommand\@@endlink[0]{}%
}{%
 \providecommand\@@startlink[1]{%
  \leavevmode
  \pdfstartlink
   attr{/Border[0 0 1 ]/H/I/C[0 1 1]}%
   user{/Subtype/Link/A<</Type/Action/S/URI/URI(#1)>>}%
  \relax
 }%
 \providecommand\@@endlink[0]{\pdfendlink}%
}%
\providecommand \url  [0]{\begingroup\@sanitize \@url }%
\providecommand \@url [1]{\endgroup\@href {#1}{\urlprefix}}%
\providecommand \urlprefix [0]{URL }%
\providecommand \Eprint[0]{\href }%
\@ifxundefined \urlstyle {%
  \providecommand \doi [1]{doi:\discretionary{}{}{}#1}%
}{%
  \providecommand \doi [0]{doi:\discretionary{}{}{}\begingroup
  \urlstyle{rm}\Url }%
}%
\providecommand \doibase [0]{http://dx.doi.org/}%
\providecommand \Doi[1]{\href{\doibase#1}}%
\providecommand \selectlanguage [0]{\@gobble}%
\providecommand \bibinfo [0]{\@secondoftwo}%
\providecommand \bibfield [0]{\@secondoftwo}%
\providecommand \translation [1]{[#1]}%
\providecommand \BibitemOpen[0]{}%
\providecommand \bibitemStop [0]{}%
\providecommand \bibitemNoStop [0]{.\EOS\space}%
\providecommand \EOS [0]{\spacefactor3000\relax}%
\providecommand \BibitemShut [1]{\csname bibitem#1\endcsname}%
%</preamble>
\bibitem{white92}%
  \BibitemOpen
  \bibfield{author}{%
  \bibinfo {author} {\bibfnamefont{S.~R.}\ \bibnamefont{White}},\ }%
  \bibfield{journal}{%
  \bibinfo {journal} {Phys. Rev. Lett}\ }%
  \textbf{\bibinfo {volume} {69}},\ \bibinfo {pages} {2863} (\bibinfo {year}
  {1992})\BibitemShut{NoStop}%
\bibitem{white92b}%
  \BibitemOpen
  \bibfield{author}{%
  \bibinfo {author} {\bibfnamefont{S.~R.}\ \bibnamefont{White}},\ }%
  \bibfield{journal}{%
  \bibinfo {journal} {Phys. Rev. B}\ }%
  \textbf{\bibinfo {volume} {48}},\ \bibinfo {pages} {10345} (\bibinfo {year}
  {1992})\BibitemShut{NoStop}%
\bibitem{xiang96}%
  \BibitemOpen
  \bibfield{author}{%
  \bibinfo {author} {\bibfnamefont{T.}~\bibnamefont{Xiang}},\ }%
  \bibfield{journal}{%
  \bibinfo {journal} {Phys. Rev. B}\ }%
  \textbf{\bibinfo {volume} {53}},\ \bibinfo {pages} {R10445} (\bibinfo {year}
  {1996})\BibitemShut{NoStop}%
\bibitem{nishimoto02}%
  \BibitemOpen
  \bibfield{author}{%
  \bibinfo {author} {\bibfnamefont{S.}~\bibnamefont{Nishimoto}}, \bibinfo
  {author} {\bibfnamefont{E.}~\bibnamefont{Jeckelmann}}, \bibinfo {author}
  {\bibfnamefont{F.}~\bibnamefont{Gebhard}},\ and\ \bibinfo {author}
  {\bibfnamefont{R.~M.}\ \bibnamefont{Noack}},\ }%
  \bibfield{journal}{%
  \bibinfo {journal} {Phys. Rev. B}\ }%
  \textbf{\bibinfo {volume} {65}},\ \bibinfo {pages} {165114} (\bibinfo {year}
  {2002})\BibitemShut{NoStop}%
\bibitem{legeza03}%
  \BibitemOpen
  \bibfield{author}{%
  \bibinfo {author} {\bibfnamefont{{\"O}.}~\bibnamefont{Legeza}}\ and\ \bibinfo
  {author} {\bibfnamefont{J.}~\bibnamefont{Solyom}},\ }%
  \bibfield{journal}{%
  \bibinfo {journal} {Phys. Rev. B}\ }%
  \textbf{\bibinfo {volume} {68}},\ \bibinfo {pages} {195116} (\bibinfo {year}
  {2003})\BibitemShut{NoStop}%
\bibitem{white99}%
  \BibitemOpen
  \bibfield{author}{%
  \bibinfo {author} {\bibfnamefont{S.~R.}\ \bibnamefont{White}}\ and\ \bibinfo
  {author} {\bibfnamefont{R.~L.}\ \bibnamefont{Martin}},\ }%
  \bibfield{journal}{%
  \bibinfo {journal} {J. Chem. Phys.}\ }%
  \textbf{\bibinfo {volume} {110}},\ \bibinfo {pages} {4127} (\bibinfo {year}
  {1999})\BibitemShut{NoStop}%
\bibitem{daul00}%
  \BibitemOpen
  \bibfield{author}{%
  \bibinfo {author} {\bibfnamefont{S.}~\bibnamefont{Daul}}, \bibinfo {author}
  {\bibfnamefont{I.}~\bibnamefont{Ciofini}}, \bibinfo {author}
  {\bibfnamefont{C.}~\bibnamefont{Daul}},\ and\ \bibinfo {author}
  {\bibfnamefont{S.~R.}\ \bibnamefont{White}},\ }%
  \bibfield{journal}{%
  \bibinfo {journal} {Int. J. Quantum Chem.}\ }%
  \textbf{\bibinfo {volume} {79}},\ \bibinfo {pages} {331} (\bibinfo {year}
  {2000})\BibitemShut{NoStop}%
\bibitem{mitrushenkov01}%
  \BibitemOpen
  \bibfield{author}{%
  \bibinfo {author} {\bibfnamefont{A.~O.}\ \bibnamefont{Mitrushenkov}},
  \bibinfo {author} {\bibfnamefont{G.}~\bibnamefont{Fano}}, \bibinfo {author}
  {\bibfnamefont{F.}~\bibnamefont{Ortolani}}, \bibinfo {author}
  {\bibfnamefont{R.}~\bibnamefont{Linguerri}},\ and\ \bibinfo {author}
  {\bibfnamefont{P.}~\bibnamefont{Palmieri}},\ }%
  \bibfield{journal}{%
  \bibinfo {journal} {J. Chem. Phys.}\ }%
  \textbf{\bibinfo {volume} {115}},\ \bibinfo {pages} {6815} (\bibinfo {year}
  {2001})\BibitemShut{NoStop}%
\bibitem{chan02}%
  \BibitemOpen
  \bibfield{author}{%
  \bibinfo {author} {\bibfnamefont{G.~K.-L.}\ \bibnamefont{Chan}}\ and\
  \bibinfo {author} {\bibfnamefont{M.}~\bibnamefont{Head-Gordon}},\ }%
  \bibfield{journal}{%
  \bibinfo {journal} {J. Chem. Phys.}\ }%
  \textbf{\bibinfo {volume} {116}},\ \bibinfo {pages} {4462} (\bibinfo {year}
  {2002})\BibitemShut{NoStop}%
\bibitem{chan03}%
  \BibitemOpen
  \bibfield{author}{%
  \bibinfo {author} {\bibfnamefont{G.~K.-L.}\ \bibnamefont{Chan}}\ and\
  \bibinfo {author} {\bibfnamefont{M.}~\bibnamefont{Head-Gordon}},\ }%
  \bibfield{journal}{%
  \bibinfo {journal} {J. Chem. Phys.}\ }%
  \textbf{\bibinfo {volume} {118}},\ \bibinfo {pages} {8551} (\bibinfo {year}
  {2003})\BibitemShut{NoStop}%
\bibitem{legeza03a}%
  \BibitemOpen
  \bibfield{author}{%
  \bibinfo {author} {\bibfnamefont{{\"O}.}~\bibnamefont{Legeza}}, \bibinfo
  {author} {\bibfnamefont{J.}~\bibnamefont{R{\"o}der}},\ and\ \bibinfo {author}
  {\bibfnamefont{B.~A.}\ \bibnamefont{Hess}},\ }%
  \bibfield{journal}{%
  \bibinfo {journal} {Phys. Rev. B}\ }%
  \textbf{\bibinfo {volume} {67}},\ \bibinfo {pages} {125114} (\bibinfo {year}
  {2003}),\
  \Eprint{http://arxiv.org/abs/cond-mat/0204602}{cond-mat/0204602}\BibitemShut%
{NoStop}%
\bibitem{legeza03b}%
  \BibitemOpen
  \bibfield{author}{%
  \bibinfo {author} {\bibfnamefont{{\"O}.}~\bibnamefont{Legeza}}, \bibinfo
  {author} {\bibfnamefont{J.}~\bibnamefont{R{\"o}der}},\ and\ \bibinfo {author}
  {\bibfnamefont{B.~A.}\ \bibnamefont{Hess}},\ }%
  \bibfield{journal}{%
  \bibinfo {journal} {Mol. Phys.}\ }%
  \textbf{\bibinfo {volume} {101}},\ \bibinfo {pages} {2019} (\bibinfo {year}
  {2003}),\
  \Eprint{http://arxiv.org/abs/cond-mat/0208187}{cond-mat/0208187}\BibitemShut%
{NoStop}%
\bibitem{moritz05}%
  \BibitemOpen
  \bibfield{author}{%
  \bibinfo {author} {\bibfnamefont{G.}~\bibnamefont{Moritz}}, \bibinfo {author}
  {\bibfnamefont{B.~A.}\ \bibnamefont{Hess}},\ and\ \bibinfo {author}
  {\bibfnamefont{M.}~\bibnamefont{Reiher}},\ }%
  \bibfield{journal}{%
  \bibinfo {journal} {J. Chem. Phys.}\ }%
  \textbf{\bibinfo {volume} {122}},\ \bibinfo {pages} {024107} (\bibinfo {year}
  {2005})\BibitemShut{NoStop}%
\bibitem{moritz06}%
  \BibitemOpen
  \bibfield{author}{%
  \bibinfo {author} {\bibfnamefont{G.}~\bibnamefont{Moritz}}, \bibinfo {author}
  {\bibfnamefont{B.~A.}\ \bibnamefont{Hess}},\ and\ \bibinfo {author}
  {\bibfnamefont{M.}~\bibnamefont{Reiher}},\ }%
  \bibfield{journal}{%
  \bibinfo {journal} {J. Chem. Phys.}\ }%
  \textbf{\bibinfo {volume} {124}},\ \bibinfo {pages} {034103} (\bibinfo {year}
  {2006})\BibitemShut{NoStop}%
\bibitem{rissler06}%
  \BibitemOpen
  \bibfield{author}{%
  \bibinfo {author} {\bibfnamefont{J.}~\bibnamefont{Rissler}}, \bibinfo
  {author} {\bibfnamefont{R.~M.}\ \bibnamefont{Noack}},\ and\ \bibinfo {author}
  {\bibfnamefont{S.~R.}\ \bibnamefont{White}},\ }%
  \bibfield{journal}{%
  \bibinfo {journal} {Chem. Phys.}\ }%
  \textbf{\bibinfo {volume} {323}},\ \bibinfo {pages} {519} (\bibinfo {year}
  {2006}),\
  \Eprint{http://arxiv.org/abs/cond-mat/0508524}{cond-mat/0508524}\BibitemShut%
{NoStop}%
\bibitem{marti08}%
  \BibitemOpen
  \bibfield{author}{%
  \bibinfo {author} {\bibfnamefont{K.}~\bibnamefont{Marti}}, \bibinfo {author}
  {\bibfnamefont{I.~M.}\ \bibnamefont{Ondik}}, \bibinfo {author}
  {\bibfnamefont{G.}~\bibnamefont{Moritz}},\ and\ \bibinfo {author}
  {\bibfnamefont{M.}~\bibnamefont{Reiher}},\ }%
  \bibfield{journal}{%
  \bibinfo {journal} {J. Chem. Phys.}\ }%
  \textbf{\bibinfo {volume} {128}},\ \bibinfo {pages} {014104} (\bibinfo {year}
  {2008})\BibitemShut{NoStop}%
\bibitem{vidallatorre03}%
  \BibitemOpen
  \bibfield{author}{%
  \bibinfo {author} {\bibfnamefont{G.}~\bibnamefont{Vidal}}, \bibinfo {author}
  {\bibfnamefont{J.~I.}\ \bibnamefont{Latorre}}, \bibinfo {author}
  {\bibfnamefont{E.}~\bibnamefont{Rico}},\ and\ \bibinfo {author}
  {\bibfnamefont{A.}~\bibnamefont{Kitaev}},\ }%
  \bibfield{journal}{%
  \bibinfo {journal} {Phys. Rev. Lett}\ }%
  \textbf{\bibinfo {volume} {90}},\ \bibinfo {pages} {227902} (\bibinfo {year}
  {2003}),\
  \Eprint{http://arxiv.org/abs/quant-ph/0211074}{quant-ph/0211074}\BibitemShut%
{NoStop}%
\bibitem{vidal03}%
  \BibitemOpen
  \bibfield{author}{%
  \bibinfo {author} {\bibfnamefont{G.}~\bibnamefont{Vidal}},\ }%
  \bibfield{journal}{%
  \bibinfo {journal} {Phys. Rev. Lett}\ }%
  \textbf{\bibinfo {volume} {91}},\ \bibinfo {pages} {147902} (\bibinfo {year}
  {2003}),\
  \Eprint{http://arxiv.org/abs/quant-ph/0310089}{quant-ph/0310089}\BibitemShut%
{NoStop}%
\bibitem{legeza04}%
  \BibitemOpen
  \bibfield{author}{%
  \bibinfo {author} {\bibfnamefont{{\"O}.}~\bibnamefont{Legeza}}\ and\ \bibinfo
  {author} {\bibfnamefont{J.}~\bibnamefont{Solyom}},\ }%
  \bibfield{journal}{%
  \bibinfo {journal} {Phys. Rev. B}\ }%
  \textbf{\bibinfo {volume} {70}},\ \bibinfo {pages} {205118} (\bibinfo {year}
  {2004})\BibitemShut{NoStop}%
\bibitem{verstraeteciracmurg08}%
  \BibitemOpen
  \bibfield{author}{%
  \bibinfo {author} {\bibfnamefont{F.}~\bibnamefont{Verstraete}}, \bibinfo
  {author} {\bibfnamefont{J.~I.}\ \bibnamefont{Cirac}},\ and\ \bibinfo {author}
  {\bibfnamefont{V.}~\bibnamefont{Murg}},\ }%
  \bibfield{journal}{%
  \bibinfo {journal} {Adv. Phys.}\ }%
  \textbf{\bibinfo {volume} {57 (2)}},\ \bibinfo {pages} {143} (\bibinfo {year}
  {2008})\BibitemShut{NoStop}%
\bibitem{verstraetecirac05}%
  \BibitemOpen
  \bibfield{author}{%
  \bibinfo {author} {\bibfnamefont{F.}~\bibnamefont{Verstraete}}\ and\ \bibinfo
  {author} {\bibfnamefont{J.}~\bibnamefont{Cirac}},\ }%
  \bibfield{journal}{%
  \bibinfo {journal} {Phys. Rev. B}\ }%
  \textbf{\bibinfo {volume} {73}},\ \bibinfo {pages} {094423} (\bibinfo {year}
  {2006}),\
  \Eprint{http://arxiv.org/abs/cond-mat/0505140}{cond-mat/0505140}\BibitemShut%
{NoStop}%
\bibitem{schuchwolf08}%
  \BibitemOpen
  \bibfield{author}{%
  \bibinfo {author} {\bibfnamefont{N.}~\bibnamefont{Schuch}}, \bibinfo {author}
  {\bibfnamefont{M.~M.}\ \bibnamefont{Wolf}}, \bibinfo {author}
  {\bibfnamefont{F.}~\bibnamefont{Verstraete}},\ and\ \bibinfo {author}
  {\bibfnamefont{J.~I.}\ \bibnamefont{Cirac}},\ }%
  \bibfield{journal}{%
  \bibinfo {journal} {Phys. Rev. Lett.}\ }%
  \textbf{\bibinfo {volume} {100}},\ \bibinfo {pages} {030504} (\bibinfo {year}
  {2008})\BibitemShut{NoStop}%
\bibitem{oestlund95}%
  \BibitemOpen
  \bibfield{author}{%
  \bibinfo {author} {\bibfnamefont{S.}~\bibnamefont{{\"O}stlund}}\ and\
  \bibinfo {author} {\bibfnamefont{S.}~\bibnamefont{Rommer}},\ }%
  \bibfield{journal}{%
  \bibinfo {journal} {Phys. Rev. Lett.}\ }%
  \textbf{\bibinfo {volume} {75}},\ \bibinfo {pages} {3537} (\bibinfo {year}
  {1995})\BibitemShut{NoStop}%
\bibitem{verstraetecirac04}%
  \BibitemOpen
  \bibfield{author}{%
  \bibinfo {author} {\bibfnamefont{F.}~\bibnamefont{Verstraete}}\ and\ \bibinfo
  {author} {\bibfnamefont{J.~I.}\ \bibnamefont{Cirac}},\ }%
  \bibfield{journal}{%
  \bibinfo {journal} {arXiv:cond-mat/0407066v1}}%
   (\bibinfo {year} {2004})\BibitemShut{NoStop}%
\bibitem{murgverstraete07}%
  \BibitemOpen
  \bibfield{author}{%
  \bibinfo {author} {\bibfnamefont{V.}~\bibnamefont{Murg}}, \bibinfo {author}
  {\bibfnamefont{F.}~\bibnamefont{Verstraete}},\ and\ \bibinfo {author}
  {\bibfnamefont{J.~I.}\ \bibnamefont{Cirac}},\ }%
  \bibfield{journal}{%
  \bibinfo {journal} {Phys. Rev. A}\ }%
  \textbf{\bibinfo {volume} {75}},\ \bibinfo {pages} {033605} (\bibinfo {year}
  {2007}),\
  \Eprint{http://arxiv.org/abs/cond-mat/0611522}{cond-mat/0611522}\BibitemShut%
{NoStop}%
\bibitem{murgverstraete08}%
  \BibitemOpen
  \bibfield{author}{%
  \bibinfo {author} {\bibfnamefont{V.}~\bibnamefont{Murg}}, \bibinfo {author}
  {\bibfnamefont{F.}~\bibnamefont{Verstraete}},\ and\ \bibinfo {author}
  {\bibfnamefont{J.~I.}\ \bibnamefont{Cirac}},\ }%
  \bibfield{journal}{%
  \bibinfo {journal} {Phys. Rev. B}\ }%
  \textbf{\bibinfo {volume} {79}},\ \bibinfo {pages} {195119} (\bibinfo {year}
  {2009})\BibitemShut{NoStop}%
\bibitem{vidal06}%
  \BibitemOpen
  \bibfield{author}{%
  \bibinfo {author} {\bibfnamefont{G.}~\bibnamefont{Vidal}},\ }%
  \bibfield{journal}{%
  \bibinfo {journal} {Phys. Rev. Lett.}\ }%
  \textbf{\bibinfo {volume} {101}},\ \bibinfo {pages} {110501} (\bibinfo {year}
  {2008})\BibitemShut{NoStop}%
\bibitem{changlani09}%
  \BibitemOpen
  \bibfield{author}{%
  \bibinfo {author} {\bibfnamefont{H.~J.}\ \bibnamefont{Changlani}}, \bibinfo
  {author} {\bibfnamefont{J.~M.}\ \bibnamefont{Kinder}}, \bibinfo {author}
  {\bibfnamefont{C.~J.}\ \bibnamefont{Umrigar}},\ and\ \bibinfo {author}
  {\bibfnamefont{G.~K.-L.}\ \bibnamefont{Chan}},\ }%
  \bibfield{journal}{%
  \bibinfo {journal} {arXiv:0907.4646v3 [cond-mat.str-el]}}%
   (\bibinfo {year} {2009})\BibitemShut{NoStop}%
\bibitem{marti10}%
  \BibitemOpen
  \bibfield{author}{%
  \bibinfo {author} {\bibfnamefont{K.~H.}\ \bibnamefont{Marti}}, \bibinfo
  {author} {\bibfnamefont{B.}~\bibnamefont{Bauer}}, \bibinfo {author}
  {\bibfnamefont{M.}~\bibnamefont{Reiher}}, \bibinfo {author}
  {\bibfnamefont{M.}~\bibnamefont{Troyer}},\ and\ \bibinfo {author}
  {\bibfnamefont{F.}~\bibnamefont{Verstraete}},\ }%
  \bibfield{journal}{%
  \bibinfo {journal} {arXiv:1004.5303v1 [physics.chem-ph]}}%
   (\bibinfo {year} {2010})\BibitemShut{NoStop}%
\bibitem{ceperley80}%
  \BibitemOpen
  \bibfield{author}{%
  \bibinfo {author} {\bibfnamefont{D.~M.}\ \bibnamefont{Ceperley}}\ and\
  \bibinfo {author} {\bibfnamefont{B.~J.}\ \bibnamefont{Alder}},\ }%
  \bibfield{journal}{%
  \bibinfo {journal} {Phys. Rev. Lett.}\ }%
  \textbf{\bibinfo {volume} {45}},\ \bibinfo {pages} {566} (\bibinfo {year}
  {1980})\BibitemShut{NoStop}%
\bibitem{troyer04}%
  \BibitemOpen
  \bibfield{author}{%
  \bibinfo {author} {\bibfnamefont{M.}~\bibnamefont{Troyer}}\ and\ \bibinfo
  {author} {\bibfnamefont{U.}~\bibnamefont{Wiese}},\ }%
  \bibfield{journal}{%
  \bibinfo {journal} {Phys.Rev.Lett.}\ }%
  \textbf{\bibinfo {volume} {94}},\ \bibinfo {pages} {170201} (\bibinfo {year}
  {2005}),\
  \Eprint{http://arxiv.org/abs/cond-mat/0408370}{cond-mat/0408370}\BibitemShut%
{NoStop}%
\bibitem{luo10}%
  \BibitemOpen
  \bibfield{author}{%
  \bibinfo {author} {\bibfnamefont{H.~G.}\ \bibnamefont{Luo}}, \bibinfo
  {author} {\bibfnamefont{M.~P.}\ \bibnamefont{Qin}},\ and\ \bibinfo {author}
  {\bibfnamefont{T.}~\bibnamefont{Xiang}},\ }%
  \bibfield{journal}{%
  \bibinfo {journal} {arXiv:1002.1287v1 [cond-mat.str-el]}}%
   (\bibinfo {year} {2010})\BibitemShut{NoStop}%
\bibitem{shi06}%
  \BibitemOpen
  \bibfield{author}{%
  \bibinfo {author} {\bibfnamefont{Y.}~\bibnamefont{Shi}}, \bibinfo {author}
  {\bibfnamefont{L.}~\bibnamefont{Duan}},\ and\ \bibinfo {author}
  {\bibfnamefont{G.}~\bibnamefont{Vidal}},\ }%
  \bibfield{journal}{%
  \bibinfo {journal} {Phys. Rev. A}\ }%
  \textbf{\bibinfo {volume} {74}},\ \bibinfo {pages} {022320} (\bibinfo {year}
  {2006})\BibitemShut{NoStop}%
\bibitem{schollwoeck04}%
  \BibitemOpen
  \bibfield{author}{%
  \bibinfo {author} {\bibfnamefont{U.}~\bibnamefont{Schollw{\"o}ck}},\ }%
  \bibfield{journal}{%
  \bibinfo {journal} {Rev. Mod. Phys.}\ }%
  \textbf{\bibinfo {volume} {77}},\ \bibinfo {pages} {259} (\bibinfo {year}
  {2005}),\
  \Eprint{http://arxiv.org/abs/cond-mat/0409292}{cond-mat/0409292}\BibitemShut%
{NoStop}%
\bibitem{cazalilla02}%
  \BibitemOpen
  \bibfield{author}{%
  \bibinfo {author} {\bibfnamefont{M.~A.}\ \bibnamefont{Cazalilla}}\ and\
  \bibinfo {author} {\bibfnamefont{J.~B.}\ \bibnamefont{Marston}},\ }%
  \bibfield{journal}{%
  \bibinfo {journal} {Phys. Rev. Lett.}\ }%
  \textbf{\bibinfo {volume} {88}},\ \bibinfo {pages} {256403} (\bibinfo {year}
  {2002})\BibitemShut{NoStop}%
\bibitem{whitefeiguin04}%
  \BibitemOpen
  \bibfield{author}{%
  \bibinfo {author} {\bibfnamefont{S.~R.}\ \bibnamefont{White}}\ and\ \bibinfo
  {author} {\bibfnamefont{A.~E.}\ \bibnamefont{Feiguin}},\ }%
  \bibfield{journal}{%
  \bibinfo {journal} {Phys. Rev. Lett}\ }%
  \textbf{\bibinfo {volume} {93}},\ \bibinfo {pages} {076401} (\bibinfo {year}
  {2004}),\
  \Eprint{http://arxiv.org/abs/cond-mat/0403310}{cond-mat/0403310}\BibitemShut%
{NoStop}%
\bibitem{daley04}%
  \BibitemOpen
  \bibfield{author}{%
  \bibinfo {author} {\bibfnamefont{A.~J.}\ \bibnamefont{Daley}}, \bibinfo
  {author} {\bibfnamefont{C.}~\bibnamefont{Kollath}}, \bibinfo {author}
  {\bibfnamefont{U.}~\bibnamefont{Schollwoeck}},\ and\ \bibinfo {author}
  {\bibfnamefont{G.}~\bibnamefont{Vidal}},\ }%
  \bibfield{journal}{%
  \bibinfo {journal} {J. Stat. Mech.:}\ }%
  \textbf{\bibinfo {volume} {Theor. Exp.}},\ \bibinfo {pages} {P04005}
  (\bibinfo {year} {2004}),\
  \Eprint{http://arxiv.org/abs/cond-mat/0403313}{cond-mat/0403313}\BibitemShut%
{NoStop}%
\bibitem{wilson75}%
  \BibitemOpen
  \bibfield{author}{%
  \bibinfo {author} {\bibfnamefont{K.~G.}\ \bibnamefont{Wilson}},\ }%
  \bibfield{journal}{%
  \bibinfo {journal} {Rev. Mod. Phys.}\ }%
  \textbf{\bibinfo {volume} {47}},\ \bibinfo {pages} {773} (\bibinfo {year}
  {1975})\BibitemShut{NoStop}%
\bibitem{hofstetter00}%
  \BibitemOpen
  \bibfield{author}{%
  \bibinfo {author} {\bibfnamefont{W.}~\bibnamefont{Hofstetter}},\ }%
  \bibfield{journal}{%
  \bibinfo {journal} {Phys. Rev. Lett.}\ }%
  \textbf{\bibinfo {volume} {85}},\ \bibinfo {pages} {1508} (\bibinfo {year}
  {2000})\BibitemShut{NoStop}%
\bibitem{peters06}%
  \BibitemOpen
  \bibfield{author}{%
  \bibinfo {author} {\bibfnamefont{R.}~\bibnamefont{Peters}}, \bibinfo {author}
  {\bibfnamefont{T.}~\bibnamefont{Pruschke}},\ and\ \bibinfo {author}
  {\bibfnamefont{F.~B.}\ \bibnamefont{Anders}},\ }%
  \bibfield{journal}{%
  \bibinfo {journal} {Phys. Rev. B}\ }%
  \textbf{\bibinfo {volume} {74}},\ \bibinfo {pages} {245114} (\bibinfo {year}
  {2006})\BibitemShut{NoStop}%
\bibitem{anders05}%
  \BibitemOpen
  \bibfield{author}{%
  \bibinfo {author} {\bibfnamefont{F.~B.}\ \bibnamefont{Anders}}\ and\ \bibinfo
  {author} {\bibfnamefont{A.}~\bibnamefont{Schiller}},\ }%
  \bibfield{journal}{%
  \bibinfo {journal} {Phys. Rev. Lett.}\ }%
  \textbf{\bibinfo {volume} {95}},\ \bibinfo {pages} {196801} (\bibinfo {year}
  {2005})\BibitemShut{NoStop}%
\bibitem{toth08}%
  \BibitemOpen
  \bibfield{author}{%
  \bibinfo {author} {\bibfnamefont{A.~I.}\ \bibnamefont{Toth}}, \bibinfo
  {author} {\bibfnamefont{C.~P.}\ \bibnamefont{Moca}}, \bibinfo {author}
  {\bibfnamefont{{\"O}.}~\bibnamefont{Legeza}},\ and\ \bibinfo {author}
  {\bibfnamefont{G.}~\bibnamefont{Zarand}},\ }%
  \bibfield{journal}{%
  \bibinfo {journal} {Phys. Rev. B}\ }%
  \textbf{\bibinfo {volume} {78}},\ \bibinfo {pages} {245109} (\bibinfo {year}
  {2008})\BibitemShut{NoStop}%
\bibitem{weichselbaum07}%
  \BibitemOpen
  \bibfield{author}{%
  \bibinfo {author} {\bibfnamefont{A.}~\bibnamefont{Weichselbaum}}\ and\
  \bibinfo {author} {\bibfnamefont{J.}~\bibnamefont{von Delft}},\ }%
  \bibfield{journal}{%
  \bibinfo {journal} {Phys. Rev. Lett.}\ }%
  \textbf{\bibinfo {volume} {99}},\ \bibinfo {pages} {076402} (\bibinfo {year}
  {2007})\BibitemShut{NoStop}%
\bibitem{holzner10}%
  \BibitemOpen
  \bibfield{author}{%
  \bibinfo {author} {\bibfnamefont{A.}~\bibnamefont{Holzner}}, \bibinfo
  {author} {\bibfnamefont{A.}~\bibnamefont{Weichselbaum}},\ and\ \bibinfo
  {author} {\bibfnamefont{J.}~\bibnamefont{von Delft}},\ }%
  \bibfield{journal}{%
  \bibinfo {journal} {Phys. Rev. B}\ }%
  \textbf{\bibinfo {volume} {81}},\ \bibinfo {pages} {125126} (\bibinfo {year}
  {2010})\BibitemShut{NoStop}%
\bibitem{georges96}%
  \BibitemOpen
  \bibfield{author}{%
  \bibinfo {author} {\bibfnamefont{A.}~\bibnamefont{Georges}}, \bibinfo
  {author} {\bibfnamefont{G.}~\bibnamefont{Kotliar}}, \bibinfo {author}
  {\bibfnamefont{W.}~\bibnamefont{Krauth}},\ and\ \bibinfo {author}
  {\bibfnamefont{M.~J.}\ \bibnamefont{Rozenberg}},\ }%
  \bibfield{journal}{%
  \bibinfo {journal} {Rev. Mod. Phys.}\ }%
  \textbf{\bibinfo {volume} {68}},\ \bibinfo {pages} {13} (\bibinfo {year}
  {1996})\BibitemShut{NoStop}%
\bibitem{eckstein05}%
  \BibitemOpen
  \bibfield{author}{%
  \bibinfo {author} {\bibfnamefont{M.}~\bibnamefont{Eckstein}}, \bibinfo
  {author} {\bibfnamefont{M.}~\bibnamefont{Kollar}}, \bibinfo {author}
  {\bibfnamefont{K.}~\bibnamefont{Byczuk}},\ and\ \bibinfo {author}
  {\bibfnamefont{D.}~\bibnamefont{Vollhardt}},\ }%
  \bibfield{journal}{%
  \bibinfo {journal} {Phys. Rev. B}\ }%
  \textbf{\bibinfo {volume} {71}},\ \bibinfo {pages} {235119} (\bibinfo {year}
  {2005})\BibitemShut{NoStop}%
\bibitem{otsuka96}%
  \BibitemOpen
  \bibfield{author}{%
  \bibinfo {author} {\bibfnamefont{H.}~\bibnamefont{Otsuka}},\ }%
  \bibfield{journal}{%
  \bibinfo {journal} {Phys. Rev. B}\ }%
  \textbf{\bibinfo {volume} {53}},\ \bibinfo {pages} {14004} (\bibinfo {year}
  {1996})\BibitemShut{NoStop}%
\bibitem{friedman97}%
  \BibitemOpen
  \bibfield{author}{%
  \bibinfo {author} {\bibfnamefont{B.}~\bibnamefont{Friedman}},\ }%
  \bibfield{journal}{%
  \bibinfo {journal} {J. Phys.: Condens. Matter}\ }%
  \textbf{\bibinfo {volume} {9}},\ \bibinfo {pages} {9021} (\bibinfo {year}
  {1997})\BibitemShut{NoStop}%
\bibitem{lepetit00}%
  \BibitemOpen
  \bibfield{author}{%
  \bibinfo {author} {\bibfnamefont{M.~B.}\ \bibnamefont{Lepetit}}, \bibinfo
  {author} {\bibfnamefont{M.}~\bibnamefont{Cousy}},\ and\ \bibinfo {author}
  {\bibfnamefont{G.}~\bibnamefont{Pastor}},\ }%
  \bibfield{journal}{%
  \bibinfo {journal} {Eur. Phys. J. B}\ }%
  \textbf{\bibinfo {volume} {13}},\ \bibinfo {pages} {421} (\bibinfo {year}
  {2000})\BibitemShut{NoStop}%
\bibitem{tagliacozzo09}%
  \BibitemOpen
  \bibfield{author}{%
  \bibinfo {author} {\bibfnamefont{L.}~\bibnamefont{Tagliacozzo}}, \bibinfo
  {author} {\bibfnamefont{G.}~\bibnamefont{Evenbly}},\ and\ \bibinfo {author}
  {\bibfnamefont{G.}~\bibnamefont{Vidal}},\ }%
  \bibfield{journal}{%
  \bibinfo {journal} {Phys. Rev. B}\ }%
  \textbf{\bibinfo {volume} {80}},\ \bibinfo {pages} {235127} (\bibinfo {year}
  {2009})\BibitemShut{NoStop}%
\bibitem{white02}%
  \BibitemOpen
  \bibfield{author}{%
  \bibinfo {author} {\bibfnamefont{S.~R.}\ \bibnamefont{White}},\ }%
  \bibfield{journal}{%
  \bibinfo {journal} {J. Chem. Phys.}\ }%
  \textbf{\bibinfo {volume} {117}},\ \bibinfo {pages} {7472} (\bibinfo {year}
  {2002})\BibitemShut{NoStop}%
\bibitem{zgid08}%
  \BibitemOpen
  \bibfield{author}{%
  \bibinfo {author} {\bibfnamefont{D.}~\bibnamefont{Zgid}}\ and\ \bibinfo
  {author} {\bibfnamefont{M.}~\bibnamefont{Nooijen}},\ }%
  \bibfield{journal}{%
  \bibinfo {journal} {J. Chem. Phys.}\ }%
  \textbf{\bibinfo {volume} {128}},\ \bibinfo {pages} {144116} (\bibinfo {year}
  {2008})\BibitemShut{NoStop}%
\bibitem{zgid08a}%
  \BibitemOpen
  \bibfield{author}{%
  \bibinfo {author} {\bibfnamefont{D.}~\bibnamefont{Zgid}}\ and\ \bibinfo
  {author} {\bibfnamefont{M.}~\bibnamefont{Nooijen}},\ }%
  \bibfield{journal}{%
  \bibinfo {journal} {J. Chem. Phys.}\ }%
  \textbf{\bibinfo {volume} {128}},\ \bibinfo {pages} {014107} (\bibinfo {year}
  {2008})\BibitemShut{NoStop}%
\bibitem{ghosh08}%
  \BibitemOpen
  \bibfield{author}{%
  \bibinfo {author} {\bibfnamefont{D.}~\bibnamefont{Ghosh}}, \bibinfo {author}
  {\bibfnamefont{J.}~\bibnamefont{Hachmann}}, \bibinfo {author}
  {\bibfnamefont{T.}~\bibnamefont{Yanai}},\ and\ \bibinfo {author}
  {\bibfnamefont{G.~K.-L.}\ \bibnamefont{Chan}},\ }%
  \bibfield{journal}{%
  \bibinfo {journal} {J. Chem. Phys.}\ }%
  \textbf{\bibinfo {volume} {128}},\ \bibinfo {pages} {144117} (\bibinfo {year}
  {2008})\BibitemShut{NoStop}%
\bibitem{molpro}%
  \BibitemOpen
  \bibfield{author}{%
  \bibinfo {author} {\bibfnamefont{R.~D.}\ \bibnamefont{Amos}}, \bibinfo
  {author} {\bibfnamefont{A.}~\bibnamefont{Bernhardsson}}, \bibinfo {author}
  {\bibfnamefont{A.}~\bibnamefont{Berning}}, \bibinfo {author}
  {\bibfnamefont{P.}~\bibnamefont{Celani}}, \bibinfo {author}
  {\bibfnamefont{D.~L.}\ \bibnamefont{Cooper}}, \bibinfo {author}
  {\bibfnamefont{M.~J.~O.}\ \bibnamefont{Deegan}}, \bibinfo {author}
  {\bibfnamefont{A.~J.}\ \bibnamefont{Dobbyn}}, \bibinfo {author}
  {\bibfnamefont{F.}~\bibnamefont{Eckert}}, \bibinfo {author}
  {\bibfnamefont{C.}~\bibnamefont{Hampel}}, \bibinfo {author}
  {\bibfnamefont{G.}~\bibnamefont{Hetzer}}, \bibinfo {author}
  {\bibfnamefont{P.~J.}\ \bibnamefont{Knowles}}, \bibinfo {author}
  {\bibfnamefont{T.}~\bibnamefont{Korona}}, \bibinfo {author}
  {\bibfnamefont{R.}~\bibnamefont{Lindh}}, \bibinfo {author}
  {\bibfnamefont{A.~W.}\ \bibnamefont{Lloyd}}, \bibinfo {author}
  {\bibfnamefont{S.~J.}\ \bibnamefont{McNicholas}}, \bibinfo {author}
  {\bibfnamefont{F.~R.}\ \bibnamefont{Manby}}, \bibinfo {author}
  {\bibfnamefont{W.}~\bibnamefont{Meyer}}, \bibinfo {author}
  {\bibfnamefont{M.~E.}\ \bibnamefont{Mura}}, \bibinfo {author}
  {\bibfnamefont{A.}~\bibnamefont{Nicklass}}, \bibinfo {author}
  {\bibfnamefont{P.}~\bibnamefont{Palmieri}}, \bibinfo {author}
  {\bibfnamefont{R.}~\bibnamefont{Pitzer}}, \bibinfo {author}
  {\bibfnamefont{G.}~\bibnamefont{Rauhut}}, \bibinfo {author}
  {\bibfnamefont{M.}~\bibnamefont{Sch{\"u}tz}}, \bibinfo {author}
  {\bibfnamefont{U.}~\bibnamefont{Schumann}}, \bibinfo {author}
  {\bibfnamefont{H.}~\bibnamefont{Stoll}}, \bibinfo {author}
  {\bibfnamefont{A.~J.}\ \bibnamefont{Stone}}, \bibinfo {author}
  {\bibfnamefont{R.}~\bibnamefont{Tarroni}}, \bibinfo {author}
  {\bibfnamefont{T.}~\bibnamefont{Thorsteinsson}},\ and\ \bibinfo {author}
  {\bibfnamefont{H.~J.}\ \bibnamefont{Werner}},\ }%
  \bibinfo {journal} {MOLPRO, a package of ab initio programs designed by H.-J.
  Werner and P. J. Knowles, version 2002.1.}\BibitemShut{Stop}%
\bibitem{yanai06}%
  \BibitemOpen
\bibfield{journal}{%
    }%
  \bibfield{author}{%
  \bibinfo {author} {\bibfnamefont{T.}~\bibnamefont{Yanai}}\ and\ \bibinfo
  {author} {\bibfnamefont{G.~K.-L.}\ \bibnamefont{Chan}},\ }%
  \bibfield{journal}{%
  \bibinfo {journal} {J. Chem. Phys.}\ }%
  \textbf{\bibinfo {volume} {124}},\ \bibinfo {pages} {194106} (\bibinfo {year}
  {2006})\BibitemShut{NoStop}%
\bibitem{yanai10}%
  \BibitemOpen
  \bibfield{author}{%
  \bibinfo {author} {\bibfnamefont{T.}~\bibnamefont{Yanai}}, \bibinfo {author}
  {\bibfnamefont{Y.}~\bibnamefont{Kurashige}}, \bibinfo {author}
  {\bibfnamefont{E.}~\bibnamefont{Neuscamman}},\ and\ \bibinfo {author}
  {\bibfnamefont{G.~K.-L.}\ \bibnamefont{Chan}},\ }%
  \bibfield{journal}{%
  \bibinfo {journal} {J. Chem. Phys.}\ }%
  \textbf{\bibinfo {volume} {132}},\ \bibinfo {pages} {024105} (\bibinfo {year}
  {2010})\BibitemShut{NoStop}%
\bibitem{moritz07}%
  \BibitemOpen
  \bibfield{author}{%
  \bibinfo {author} {\bibfnamefont{G.}~\bibnamefont{Moritz}}\ and\ \bibinfo
  {author} {\bibfnamefont{M.}~\bibnamefont{Reiher}},\ }%
  \bibfield{journal}{%
  \bibinfo {journal} {J. Chem. Phys.}\ }%
  \textbf{\bibinfo {volume} {126}},\ \bibinfo {pages} {244109} (\bibinfo {year}
  {2007})\BibitemShut{NoStop}%
\end{thebibliography}

%

\end{document}